\documentclass[preprint,12pt]{elsarticle}

\usepackage{hyperref}
\usepackage{graphicx}
\journal{Computers and Fluids}  

\begin{document}

\begin{frontmatter}

\title{Nonreflecting outlet boundary conditions for incompressible flows
using SPH}



\author[address1]{Carlos E. Alvarado-Rodr\'{\i}guez}
\ead{iqcarlosug@gmail.com}

\author[address2,address3]{Jaime Klapp\corref{mycorrespondingauthor}}
\cortext[mycorrespondingauthor]{Corresponding author}
\ead{jaime.klapp@inin.gob.mx}

\author[address4]{Leonardo Di G. Sigalotti}
\ead{leonardo.sigalotti@gmail.com}

\author[address5]{Jos\'e M. Dom\'{\i}nguez}
\ead{jmdominguez@uvigo.es}

\author[address2]{Eduardo de la Cruz S\'anchez}
\ead{eduardo.delacruz@inin.gob.mx}

\address[address1]{Departamento de Ingenier\'{\i}a Qu\'{\i}mica, DCNyE, Universidad
de Guanajuato, Noria Alta S/N, 36000 Guanajuato, Guanajuato, Mexico}

\address[address2]{Departamento de F\'{\i}sica, Instituto Nacional de Investigaciones
Nucleares (ININ), Carretera M\'exico-Toluca S/N, La Marquesa, 52750 Ocoyoacac, Estado de 
M\'exico, Mexico}

\address[address3]{ABACUS-Centro de Matem\'aticas Aplicadas y C\'omputo de Alto
Rendimiento, Departamento de Matem\'aticas, Centro de Investigaci\'on y de Estudios
Avanzados (Cinvestav-IPN), Carretera M\'exico-Toluca km. 38.5, La Marquesa, 52740
Ocoyoacac, Estado de M\'exico, Mexico}

\address[address4]{\'Area de F\'{\i}sica de Procesos Irreversibles, Departamento de
Ciencias B\'asicas, Universidad Aut\'onoma Metropolitana-Azcapotzalco (UAM-A),
Av. San Pablo 180, 02200 M\'exico D.F., Mexico}

\address[address5]{EPHYSLAB (Environmental Physics Laboratory), Facultad de Ciencias,
Campus de Ourense, Universidad de Vigo, 32004 Ourense, Spain}

\begin{abstract}
In this paper we implement a simple strategy, based on Jin and Braza's method, to deal 
with nonreflecting outlet boundary conditions for incompressible Navier-Stokes flows
using the method of smoothed particle hydrodynamics (SPH). The outflow boundary 
conditions are implemented using an outflow zone downstream of the outlet, where particles
are moved using an outgoing wave equation for the velocity field so that feedback noise
from the outlet boundary is greatly reduced. For unidirectional flow across the outlet,
this condition reduces to Orlanski's wave equation. The performance of the method is
demonstrated through several two-dimensional test problems, including unsteady, 
plane Poiseuille flow, flow between two inclined plates, the Kelvin-Helmholtz instability
in a channel, and flow in a constricted conduit, and in three-dimensions for turbulent flow
in a $90^{\circ}$ section of a curved square pipe. The results show that spurious waves
incident from the outlet are effectively absorbed and that steady-state laminar flows
can be maintained for much longer times compared to periodic boundary conditions. In
addition, time-dependent anisotropies in the flow, like fluid recirculations, are
convected across the outlet in a very stable and accurate manner.
\end{abstract}


\end{frontmatter}

%\linenumbers

\section{Introduction}

A standard procedure in engineering fluid dynamics for simulating internal flows
is to artificially truncate their actual physical domain into a short region to
reduce the computational cost. However, this demands the use of one or more open
boundary conditions that must be specified at the extremes of the computational
domain. Typical examples of this type of flows include pipe and channel flows,
constricted flows in a pipe section, flows around bluff bodies, and external flows,
in which case the flow develops in a practically infinite free stream, as in rising
fire and hydrothermal seawater plumes, among others. Such open boundary conditions
must allow the fluid to enter (inlet) and leave (outlet) the computational domain 
while reasonably minimizing non-physical feedbacks \cite{Sani94}, such as the
artificial build-up of fluid near the outlet \cite{Lastiwka09} and the reflection
of outgoing waves in recirculating flows at high and moderate Reynolds numbers 
\cite{Dong06}. In addition, numerical instabilities may well develop when strong
vortices are convected through the outlet \cite{Dong14}.

In this paper, we restrict ourselves to weakly compressible flows and introduce 
a practical methodology for handling nonreflecting outlet boundary conditions with
smoothed particle hydrodynamics (SPH), when flow anisotropies are present near
the outlet. Because of its Lagrangian character, SPH presents inherent disadvantages
in the treatment of open boundary conditions compared to traditional Eulerian methods,
where inlet and outlet boundary conditions for a stationary flow are more
naturally described \cite{Lastiwka09}. Earlier attempts of handling inlet/outlet 
boundaries with SPH for simple flows, such as Poiseuille and Couette flows, have
been performed using periodic boundary conditions \cite{Morris97}. In this case,
the particle distribution is continually recycled so that any time a particle 
crosses the outlet, it is forced to re-enter the domain through the inlet.
Although this interpolation has been widely employed in SPH simulations of internal,
incompressible flows \cite{Lee08,Shadloo11,Vazquez12}, it is known to degrade 
the numerical solution over time because disturbances in the particle distribution
are re-introduced into the computational domain. For steady-state flows this
problem is mitigated by just placing the outlet plane sufficiently far from the
inlet to allow the numerical oscillations to be dissipated \cite{Khorasanizade12a}. 
However, this may incur in an excessive increase of the computational burden.

A number of approaches has been presented in the literature to prescribe the 
correct inlet and outlet boundary conditions for weakly compressible and
incompressible flows in SPH, with all of them aimed at reducing the level of
disturbance at the outflow and therefore the reflections that may
significantly alter the flow upstream. Truly incompressible SPH schemes (ISPH)
employ a pressure-correction projection scheme to compute the pressure from a
Poisson equation, which is then used to make the velocity field divergence-free 
\cite{Chorin68,Temam69}. With these methods, solution of the Poisson equation
typically requires implementing a homogeneous Neumann condition for the pressure
\cite{Lee08,Cummins99}. An improved algorithm, based on a non-homogeneous pressure
boundary condition, the so-called rotational pressure-correction scheme
\cite{Timmermans96}, was implemented in ISPH by Hosseini and Feng \cite{Hosseini11}.
Other strategies based on the use of a time-dependent driving force 
\cite{Khorasanizade12b} and the influx of kinetic energy into the domain through
the outflow boundaries \cite{Dong14} have also been designed.

On the other hand, weakly compressible SPH (WCSPH) solves the Navier-Stokes 
equations by defining the pressure through an algebraic relation so that the sound
speed is artificially set to achieve accurate results in fluid propagation \cite{Becker07}.
It is common practice in SPH to deal with spatially fixed (Eulerian) inlet and outlet
boundaries by defining inflow and outflow regions that are external to the computational
domain \cite{Morris97,Lastiwka09,Khorasanizade12a,Liang14}. These regions are filled
with particles and their widths are comparable to or greater than the smoothing length
of the fluid particles to avoid truncation of the kernel function. Thus, as a particle
pertaining to the inflow region enters the fluid domain, a new one is automatically
created to compensate it. According to the flow rates across the inlet and outlet,
inflow and outflow particles are being added and removed, respectively. Here we
describe a methodology, based on inflow and outflow zone particles, that conserves
the global mass of the system and minimizes the reflection of disturbances from the
outlet back into the fluid domain. To this end, the velocity vector of particles
in the outflow zone is evolved by means of an anisotropic propagation wave equation
following the procedure described by Jin and Braza \cite{Jin93}. In addition to
adapting their procedure for use in SPH simulations, the method is extended to three-space
dimensions. We validate the method against several benchmark test cases for the
simulation of two-dimensional (2D) internal flows, including unsteady, plane
Poiseuille flow, flow along a divergent duct, the Kelvin-Helmholtz instability of
flow with discontinuous shear in a channel, and plane choked flow through a constricted 
conduit. Validation of the method in three dimensions
(3D) is shown against physical experiments for the turbulent flow in a $90^{\circ}$
section of a curved square pipe.

\section{WCSPH Formulation}

For viscous incompressible flows, the governing equations are given by the Navier-Stokes
equations
\begin{equation}
\frac{d{\bf v}}{dt}=-\frac{1}{\rho}\nabla p+\nu\nabla ^{2}{\bf v},
\end{equation}
where $\rho$ is the density, $p$ the pressure, ${\bf v}$ the velocity field, and $\nu$
the kinematic viscosity. In a WCSPH formulation, where the pressure is given as a
function of the density, local variations of the pressure gradient may induce local
density fluctuations in the flow. Therefore, the flow is modelled by an artificial fluid
that is approximately incompressible. This is done by defining the total pressure gradient
in Eq. (1) as \cite{Morris97}
\begin{equation}
-\frac{1}{\rho}\nabla p=-\frac{1}{\rho}\nabla p_{d}-\frac{1}{\rho}\nabla p_{h}=
-\frac{1}{\rho}\nabla p_{d}+{\bf F},
\end{equation}
where $p_{d}$ is the dynamical pressure as calculated from the equation of state, $p_{h}$
is the hydrostatic pressure, and the term $-\nabla p_{h}/\rho$ is treated as a body force
${\bf F}$ to be determined. In a WCSPH scheme, the mass of a fluid element remains
constant and only its associated density fluctuates. Such density fluctuations are
calculated by solving the continuity equation
\begin{equation}
\frac{d\rho}{dt}=-\rho\nabla\cdot {\bf v}.
\end{equation}
The dynamical pressure $p_{d}$, which for simplicity we shall denote by $p$, is calculated
using the relation \cite{Becker07}
\begin{equation}
p=p_{0}\left[\left(\frac{\rho}{\rho _{0}}\right)^{\gamma}-1\right],
\end{equation}
where $\gamma =7$, $p_{0}=c_{0}^{2}\rho _{0}/\gamma$, $\rho _{0}$ is a 
reference density, and $c_{0}$ is the sound speed at the reference density. This equation
enforces very low density fluctuations since the speed of sound can be artificially
slowed with accurate results in fluid propagation. By restricting the sound speed to be
at least 10 times higher than the maximum expected fluid velocity, the density fluctuations
will be within 1\%.

In order to capture coherent turbulent structures within turbulent flows, the standard
SPH viscous formulation is replaced by a sub-particle scaling (SPS) technique \cite{Dalrymple06}.
This is achieved by Favre-averaging Eqs. (1) and (3) over a length scale comparable to
the particle sizes, where the velocity field ${\bf v}$ can be decomposed into a mean
part $\tilde {\bf v}$ and a fluctuating part ${\bf v}^{\prime}$ such that
${\bf v}=\tilde {\bf v}+{\bf v}^{\prime}$, where the mean part is defined by a density weighted 
avarage, $\tilde {\bf v}=\overline{\rho {\bf v}}/{\overline\rho}$, and the overbars denote an
arbitrary spatial filtering. Applying a flat-top spatial-filter to Eqs. (1) and (3), they become
\cite{Yoshizawa86}
\begin{eqnarray}
\frac{d\tilde {\bf v}}{dt}&=&-\frac{1}{\overline\rho}\nabla {\overline p}+
\frac{\nu}{\overline\rho}
\left[\nabla\cdot\left(\overline\rho\nabla\right)\right]{\tilde {\bf v}}+
\frac{\nu}{\overline\rho}\nabla\cdot {\bf T},\\
\frac{d\overline\rho}{dt}&=&-{\overline\rho}\nabla\cdot\tilde {\bf v},
\end{eqnarray}
respectively, where ${\bf T}$ is the SPS stress tensor defined in component form as
\begin{equation}
T_{ij}=\overline\rho\left(2\nu _{t}{\tilde S}_{ij}-\frac{2}{3}{\tilde S}_{kk}\delta _{ij}
\right)-\frac{2}{3}\overline\rho C_{I}\Delta ^{2}\delta _{ij},
\end{equation}
where
\begin{equation}
{\tilde S}_{ij}=-\frac{1}{2}\left(\frac{\partial {\tilde v}_{i}}{\partial x_{j}}+
\frac{\partial {\tilde v}_{j}}{\partial x_{i}}\right),
\end{equation}
is the Favre-averaged strain rate tensor, $C_{I}=0.00066$, 
$\nu _{t}=(C_{s}\Delta)^{2}|\tilde S|$, with $C_{s}=0.12$, is the Smagorinsky eddy
viscosity, $|\tilde S|=(2{\tilde S}_{ij}{\tilde S}_{ij})^{1/2}$ is the local strain
rate, $\delta _{ij}$ is the Kronecker delta, and $\Delta$ is a measure of the initial
particle spacing.

\section{SPH solver}

The computer code used in this work is based on standard SPH methods
\cite{Monaghan92,Libersky93}, where the density of particle $a$ is given by the usual
kernel summation
\begin{equation}
\rho _{a}=\sum _{b=1}^{n}m_{b}W_{ab}.
\end{equation}
In this expression, $m_{b}$ is the mass of particle $b$,
$W_{ab}=W(|{\bf x}_{a}-{\bf x}_{b}|,h)$ is the kernel function, where 
${\bf x}_{a}-{\bf x}_{b}$ is the distance between particles $a$ and $b$ and $h$ is
the width of the kernel or smoothing length, and the summation is taken over all $n$
neighbour particles within the kernel support. Note that the density in Eq. (9) may
be either the local density $\rho$ or the particle-scale density $\overline\rho$ 
depending on whether
we are dealing with laminar or turbulent (rotational) flows. The same is true for
the velocity field, where ${\bf v}$ may represent a local value (for laminar
flows) or the Favre-averaged velocity $\tilde {\bf v}$ (for turbulent flows).

In Eqs. (1) and (5) the pressure gradient is written in SPH form using the symmetric
representation proposed by Colagrossi and Landrini \cite{Colagrossi03}, which ensures
numerical stability at the interface between two media with large density differences,
while the laminar viscous term and the SPS stresses are discretized according to the
formulations given by Lo and Shao \cite{Lo02}. Therefore, in SPH form Eq. (5) reads
\begin{eqnarray}
\frac{d{\bf v}_{a}}{dt}=&-&\frac{1}{\rho _{a}}\sum _{b=1}^{n}\frac{m_{b}}{\rho _{b}}
\left(p_{a}+p_{b}\right)\nabla _{a}W_{ab}+4\nu\sum _{b=1}^{n}m_{b}
\frac{{\bf v}_{a}-{\bf v}_{b}}{\rho _{a}+\rho _{b}}\frac{{\bf x}_{ab}\cdot\nabla _{a}W_{ab}}
{|{\bf x}_{ab}|^{2}+\epsilon ^{2}}\nonumber\\
&+&\sum _{b=1}^{n}m_{b}\left(\frac{{\bf T}_{a}}{\rho _{a}^{2}}+
\frac{{\bf T}_{b}}{\rho _{b}^{2}}\right)\cdot\nabla _{a}W_{ab},
\end{eqnarray}
where ${\bf x}_{ab}={\bf x}_{a}-{\bf x}_{b}$, $\epsilon ^{2}=0.01h^{2}$, and $W_{ab}$ is
evaluated according to the symmetrized kernel function \cite{Hernquist89}
\begin{equation}
W_{ab}=\frac{1}{2}\left[W(|{\bf x}_{a}-{\bf x}_{b}|,h_{a})+
W(|{\bf x}_{a}-{\bf x}_{b}|,h_{b})\right],
\end{equation}
which has the correct limiting behaviour when $h_{a}=h_{b}$. For laminar flows the SPH
representation of Eq. (1) can be recovered from the discrete Eq. (10) by dropping the
SPS stress term and keeping in mind that the fluid variables will now correspond to local
quantities. Coupled to Eqs. (9) and (10), the equation
\begin{equation}
\frac{d{\bf x}_{a}}{dt}={\bf v}_{a}+\frac{\beta}{M}\sum _{b=1}^{N}m_{b}
\frac{{\bf x}_{ab}}{\left({\bf x}_{ab}\cdot {\bf x}_{ab}\right)^{3/2}}
x_{0}v_{\rm max},
\end{equation}
is solved for the particle positions, where the second term on
the right-hand side is the shifting vector of particle $a$ \cite{Vacondio13},
which modifies the position of particles in order to prevent magnification
of the SPH discretization errors due to anisotropies in their
distribution. Here, $\beta$ is a dimensionless parameter which is chosen
to be $\beta =0.04$, $v_{\rm max}$ is the maximum velocity in the system,
$M$ is the total mass
\begin{equation}
M=\sum _{b=1}^{N}m_{b},
\end{equation}
and 
\begin{equation}
x_{0}=\frac{1}{N}\sum _{b=1}^{N}\left({\bf x}_{ab}\cdot {\bf x}_{ab}\right)^{1/2}.
\end{equation}
Note that the summations in the above two expressions are over all particles
filling the fluid domain. The addition of the shifting vector on
the right-hand side of Eq. (12) does not affect momentum preservation.

Since direct evaluation of second-order derivatives of the kernel is not
required, we adopt a low-order, Wendland C$^{2}$ function \cite{Dehnen12} as
the interpolation kernel
\begin{equation}
W(q,h)=\frac{7}{4\pi h^{2}}\left(1-\frac{q}{2}\right)^{4}
\left(2q+1\right),
\end{equation}
for $0\leq q<2$ and zero otherwise, where $q=|{\bf x}-{\bf x}^{\prime}|/h$.
A Verlet algorithm is used for the time integration of Eqs. (10) and (12),
where the velocities and positions of particles are advanced from time
$t^{n}$ to time $t^{n+1}=t^{n}+\Delta t$ according to the difference
formulae
\begin{eqnarray}
{\bf v}_{a}^{n+1}&=&{\bf v}_{a}^{n-1}+2\Delta t\left(\frac{d{\bf v}_{a}}{dt}
\right)^{n},\nonumber\\
{\bf x}_{a}^{n+1}&=&{\bf x}_{a}^{n}+\Delta t{\bf v}_{a}^{n}+0.5\Delta t^{2}
\left(\frac{d{\bf v}_{a}}{dt}\right)^{n}.
\end{eqnarray}
In order to improve the coupling of Eqs. (10) and (12) during the entire
evolution, the above time integration is replaced every 50 time steps by the
alternative difference forms
\begin{eqnarray}
{\bf v}_{a}^{n+1}&=&{\bf v}_{a}^{n}+\Delta t\left(\frac{d{\bf v}_{a}}{dt}
\right)^{n},\nonumber\\
{\bf x}_{a}^{n+1}&=&{\bf x}_{a}^{n}+\Delta t{\bf v}_{a}^{n}+0.5\Delta t^{2}
\left(\frac{d{\bf v}_{a}}{dt}\right)^{n}.
\end{eqnarray}
This prevents the time integration to produce results that diverge from the
actual solution. The time step, $\Delta t$, is calculated using the
Courant-Friedrichs-Lewy (CFL) and the viscous diffusion conditions such
that
\begin{eqnarray}
\Delta t_{f,a}&=&\min _{a}\left(h|dv_{a}/dt|^{-1}\right)^{1/2},\nonumber\\
\Delta t_{v,a}&=&\max _{b}|h{\bf x}_{ab}\cdot{\bf v}_{ab}/({\bf x}_{ab}
\cdot {\bf x}_{ab}+\epsilon ^{2})|,\nonumber\\
\Delta t_{cv,a}&=&\min _{a}\left[h\left(c_{a}+\Delta t_{v,a}\right)^{-1}
\right],\nonumber\\
\Delta t&=&0.3\min _{a}\left(\Delta t_{f,a},\Delta t_{cv,a}\right),
\end{eqnarray}
where the maximum and minima are taken over all particles in the system,
$v_{a}=({\bf v}_{ab}\cdot{\bf v}_{ab})^{1/2}$, and $c_{a}$ is the
sound speed for particle $a$.

\subsection{Solid boundary conditions}

No-slip boundary conditions are implemented at contact with solid surfaces. A
stable and accurate approach is achieved here using the method of image
particles \cite{Morris97}, where imaginary particles are initially created by
simply reflecting actual fluid particles across the solid surface. Such
particles are external to the fluid domain and serve to remove the kernel
truncation in the proximity of the surface. Unlike actual fluid
particles, imaginary particles are not allowed to move relative to the solid
surface and are forced to maintain their initial distribution during the time 
evolution. However, a velocity needs be assigned to each imaginary particle in
order to evaluate the compressional and viscous forces in Eq. (10).

\section{Inlet and outlet boundary conditions}

We consider flow through a truncated section of a pipe, or channel, and assume
that the open boundaries at the entrance and exit of the pipe section are
perfectly planar. Since SPH particles cannot be spatially fixed at the planar
boundaries, we must allow them to flow in and out consistently with the flow
rate across the inlet and outlet planes, respectively.

The method as presented here distinguishes among three zones, which
are external to the computational domain, i.e., an inflow zone, which is placed
in front of the pipe inlet, an outflow zone, which is placed downstream of the pipe
exit plane, and a reservoir zone, where inert particles are temporarily stored.
The inflow region consists of five columns of uniformly spaced particles and has
a length equal to $5\Delta x_{0}$, where $\Delta x_{0}$ is the initial uniform
spacing of the fluid particles in the direction of the flow. As in Refs.
\cite{Morris97,Lastiwka09,Hosseini11,Liang14}, inflow particles are allowed to
cross the inlet plane with a prescribed velocity that may vary in time and/or
space. Scalar variables, such as density and pressure, are also prescribed for
inflow particles. Since particles close to the inlet, but inside the flow
domain, are updated according to Eqs. (9), (10), and (12), it always happen that
some inflow particles that are close to the inlet fall within the kernel support
of the near-boundary fluid particles. This will allow boundary information to be
propagated into the flow domain \cite{Lastiwka09}.

The main differences between this and previously reported methods for WCSPH flows in
SPH lie on the treatment of the outflow zone particles and the use of a reservoir
buffer to ensure conservation of both the total mass and the total number of particles.
The length of the outflow zone is chosen to be the same of the inflow zone. It is
common practice to put this boundary sufficiently far from any sources of flow
anisotropy, as may be the case of flow past a backward facing step, where the
flow becomes essentially unidirectional and approaches a steady-state regime.
In this case, classical ``do-nothing'' conditions \cite{Gresho91,Glowinski92},
where $d{\bf v}/dt=0$ and
\begin{equation}
{\bf T}\cdot {\bf n}=0,
\end{equation}
at the outlet, have become the most widely used outflow conditions for the
Navier-Stokes equations. Since these conditions are strictly valid for steady-state, 
fully-developed flows, they may present the problem of upstream wave propagation
from the outlet if anisotropies are being convected into the outflow zone. An
extension of the ``do-nothing'' conditions that enhances the stability properties
against non-physical feedbacks has recently been proposed by Braack and Mucha
\cite{Braack14}.

Here we implement a type of outflow boundary condition that simulates the
propagation of waves out of the computational domain by allowing the flow to
cross the outlet without being significantly reflected back. To do so we adopt the
procedure described in Ref. \cite{Jin93}, which is based on a wave equation and
allows for anisotropic wave propagation across the outlet. The velocity vector
of particles crossing the outlet and entering the outflow zone is considered
as a transported wave quantity incident on the boundary. To this end, we consider
the wave equation
\begin{equation}
\frac{\partial ^{2}{\bf v}}{\partial t^{2}}-c_{x}^{2}\frac{\partial ^{2}{\bf v}}
{\partial x^{2}}-c_{y}^{2}\frac{\partial ^{2}{\bf v}}{\partial y^{2}}-c_{z}^{2}
\frac{\partial ^{2}{\bf v}}{\partial z^{2}}={\bf 0},
\end{equation}
where $c_{x}$, $c_{y}$, and $c_{z}$ are the characteristic velocities of wave propagation
in the $x$-, $y$-, and $z$-directions, respectively. Introducing the differential 
operator ${\cal L}$ as
\begin{equation}
{\cal L}=c_{x}^{2}\frac{\partial ^{2}}{\partial x^{2}}+c_{y}^{2}\frac{\partial ^{2}}
{\partial y^{2}}+c_{z}^{2}\frac{\partial ^{2}}{\partial z^{2}}-\frac{\partial ^{2}}
{\partial t^{2}},
\end{equation}
Eq. (20) can be rewritten as
\begin{equation}
{\cal L}{\bf v}={\cal L}^{+}{\cal L}^{-}{\bf v}={\bf 0},
\end{equation}
where ${\cal L}^{+}$ and ${\cal L}^{-}$ are factorization operators providing
information on the outgoing and ingoing (reflected) waves, respectively. The decomposition
of ${\cal L}$ into the product ${\cal L}^{+}{\cal L}^{-}$ yields the forms
\begin{eqnarray}
{\cal L}^{+}&=&c_{x}\frac{\partial}{\partial x}+\frac{\partial}{\partial t}
\left(1-s^{2}\right)^{1/2},\\
{\cal L}^{-}&=&c_{x}\frac{\partial}{\partial x}-\frac{\partial}{\partial t}
\left(1-s^{2}\right)^{1/2},
\end{eqnarray}
where 
\begin{equation}
s^{2}=c^{2}_{y}\left(\frac{\partial}{\partial y}\right)^{2}\left(\frac{\partial}{\partial t}
\right)^{-2}+c^{2}_{z}\left(\frac{\partial}{\partial z}\right)^{2}
\left(\frac{\partial}{\partial t}\right)^{-2}.
\end{equation}
Application of the equation ${\cal L}^{-}{\bf v}={\bf 0}$ to the outflow particles
results in a total non-reflecting condition. Now, using the approximation
$(1-s^{2})^{1/2}\approx 1-s^{2}/2$ for $s$ small and making 
$c_{y}\approx c_{z}=c$, gives for the outgoing wave
\begin{equation}
\frac{\partial {\bf v}}{\partial t}+c_{x}\frac{\partial {\bf v}}{\partial x}
-\frac{1}{2}c^{2}\left(\frac{\partial}{\partial t}\right)^{-1}\left(
\frac{\partial ^{2}{\bf v}}{\partial y^{2}}+\frac{\partial ^{2}{\bf v}}
{\partial z^{2}}\right)={\bf 0},
\end{equation}
where the anisotropic term containing the coefficient $c$ is a diffusion-like
term. Noting that $c^{2}(\partial /\partial t)^{-1}$ has the same dimensions
of the kinematic viscosity $\nu$, a matching of Eq. (26) with the Navier-Stokes 
equations can be made by applying 
the following equivalences $2\nu\to c^{2}(\partial /\partial t)^{-1}$ and
$c_{x}\to v_{x}$, where $v_{x}$ is the $x$-component of the velocity field. In
this way, the outgoing wave equation becomes
\begin{equation}
\frac{\partial {\bf v}}{\partial t}+v_{x}\frac{\partial {\bf v}}{\partial x}
-\nu\left(\frac{\partial ^{2}{\bf v}}{\partial y^{2}}+
\frac{\partial ^{2}{\bf v}}{\partial z^{2}}\right)={\bf 0},
\end{equation}
where ${\bf v}=(v_{x},v_{y},v_{z})$.
If the diffusion term is dropped, Eq. (27) reduces to
\begin{equation}
\frac{\partial {\bf v}}{\partial t}+v_{x}\frac{\partial {\bf v}}{\partial x}={\bf 0},
\end{equation}
which is the Orlanski equation for unidirectional monochromatic travelling waves
\cite{Sohankar98}. Note that setting ${\bf v}=(v_{x},v_{y})$ with $v_{x}$ and
$v_{y}$ depending only on $x$ and $y$, Eq. (27) reduces to the 2D form derived by
Jin and Braza \cite{Jin93}.

\begin{figure}
\includegraphics[width=10cm]{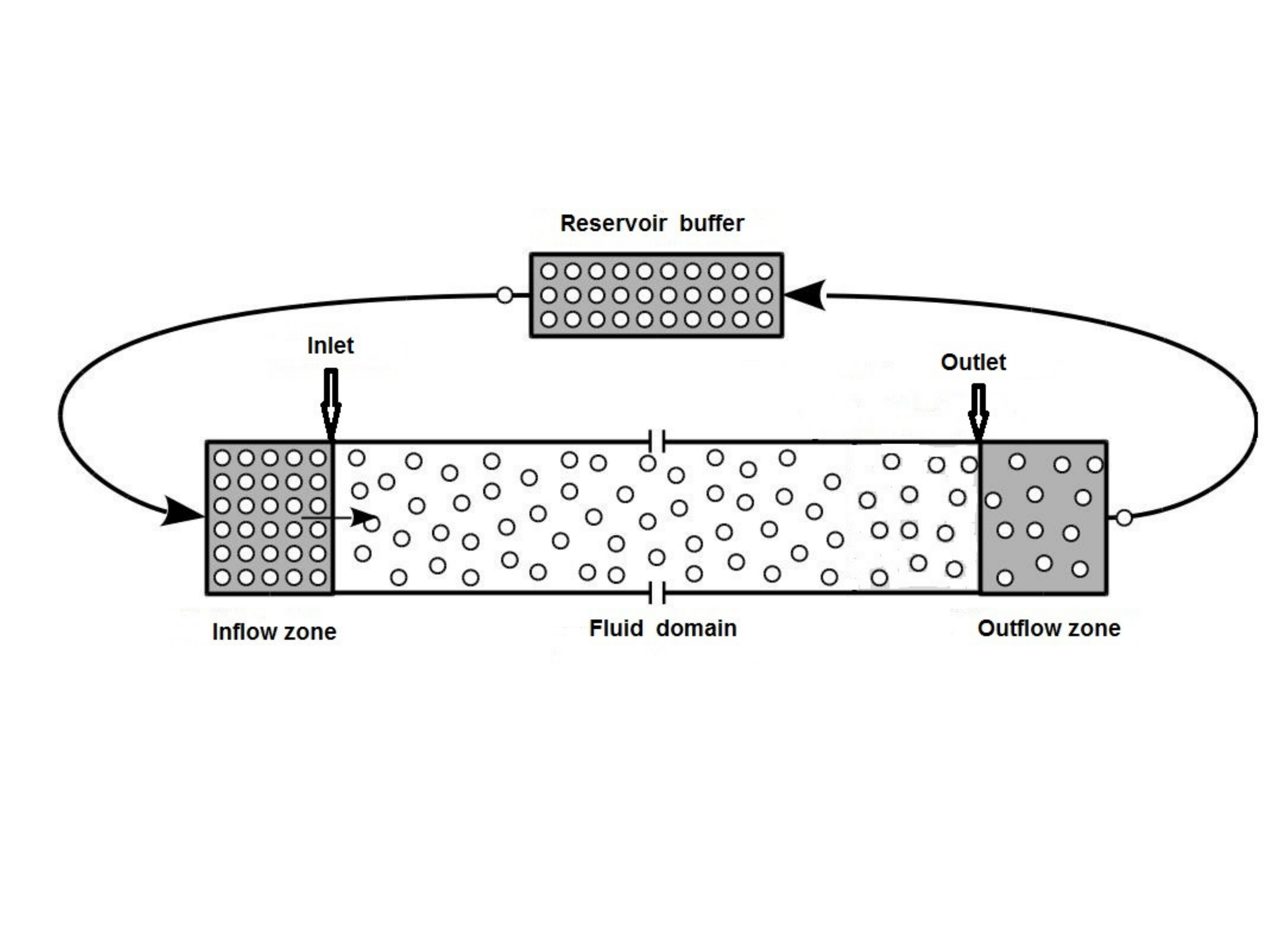}
\centering
\caption{Schematic drawing of the inflow, outflow, and reservoir zones. When an
inflow particle enters the fluid domain, a reservoir particle is automatically
inserted in the first layer of inflow particles. Similarly, when a particle
leaves the outflow zone it is temporarily stored in the reservoir buffer.}
\end{figure}

Particles in the outflow zone are evolved using either Eq. (27) or (28) until
they flow past its downstream limit. When a particle leaves the outflow zone its
velocity is automatically zeroed and it is temporarily stored in a reservoir
buffer. Figure 1 shows a schematic diagram of the inflow and outflow boundary
zones. Every time an inflow particle enters the fluid domain, a particle is  
removed from the reservoir buffer and inserted in the upstream side of the
inflow zone with the desired prescribed velocity and density. This is a
necessary step because in most problems of interest the inlet and outlet mass
rates and cross-sections may differ. At the beginning of a calculation the
number of reservoir particles depends on the flow model and can be as large as
needed.

In SPH form Eq. (27) for an outflow zone particle ``o'' is written as follows
\begin{eqnarray}
\frac{\partial {\bf v}_{o}}{\partial t}&=&-v_{x,o}\sum _{b=1}^{n}\frac{m_{b}}
{{\bar\rho}_{ob}}\left({\bf v}_{b}-{\bf v}_{o}\right)\frac{\partial W_{ob}}
{\partial x_{o}}\nonumber\\
&+&2\nu\sum _{b=1}^{n}\frac{m_{b}}{\rho _{b}}
\frac{\left({\bf v}_{b}-{\bf v}_{o}\right)}{|{\bf x}_{ob}|^{2}+\epsilon ^{2}}
\left(y_{ob}\frac{\partial W_{ob}}{\partial y_{o}}+z_{ob}
\frac{\partial W_{ob}}{\partial z_{o}}\right),
\end{eqnarray} 
where ${\bf x}_{ob}={\bf x}_{o}-{\bf x}_{b}$, $y_{ob}=y_{o}-y_{b}$, and
$z_{ob}=z_{o}-z_{b}$. Flow particles next to the outlet
plane are updated according to the usual SPH procedures so that some outflow
particles may fall inside the compact support of the near-boundary fluid particles.
The same is true for outflow particles close to the outlet in Eq. (29) where
some neighbours $b$ may actually be fluid particles, allowing fluid information
to be propagated into the outflow zone. In order to ensure stability of
Eq. (29), the velocity $v_{x,o}$ is smoothed according to
\begin{equation}
v_{x,o}=\sum _{b=1}^{n}\frac{m_{b}}{\rho _{b}}v_{x,b}W_{ob},
\end{equation}
where the summation is taken over all neighbours of outflow zone particle ``o''.
This is, in fact, equivalent to averaging the convective velocity at the outlet.
The position of outflow particles is
updated according to $d{\bf x}_{o}/dt={\bf v}_{o}$, which together
with Eq. (29), is integrated in time using the Verlet algorithm described
by Eqs. (16) and (17).

\section{Numerical tests}

\subsection{Unsteady plane Poiseuille flow}

\begin{figure}
\centering
\includegraphics[width=10cm]{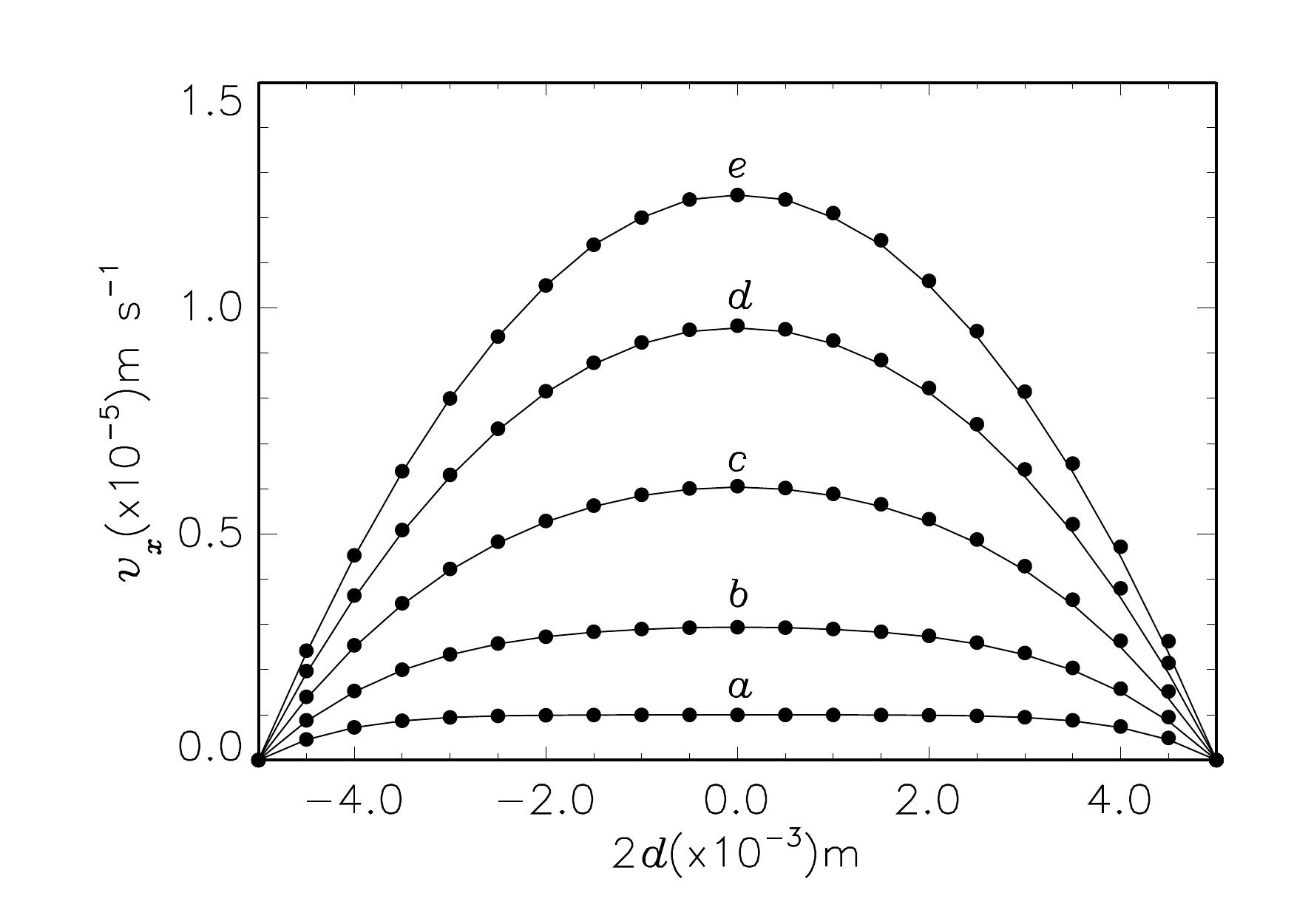}
\caption{SPH velocity profiles using the nonreflecting outlet boundary condition
(dots) as compared with the analytical
solution (solid lines) for unsteady plane Poiseuille flow with Reynolds
number, Re$=0.0125$, at: (a) 0.01 s, (b) 0.03 s, (c)  0.07 s, (d) 0.15 s, and
(e) 1.0 s. A steady-state profile is already obtained at about 1.0 s.}
\end{figure}

As a first test we consider the Poiseuille flow between stationary infinite
plates placed at distances $y=\pm 5\times 10^{-4}$ m from the centre $y=0$.
For this test, the Navier-Stokes equations admit the exact solution
\begin{eqnarray}
v_{x}(y,t)&=&\frac{F}{2\nu}\left(y^{2}-d^{2}\right)\nonumber\\
&+&\sum _{n=0}^{\infty}
\frac{16(-1)^{n}d^{2}F}{\nu\pi ^{3}(2n+1)^{3}}\cos\left[
\frac{(2n+1)\pi y}{2d}\right]\exp \left[-\frac{(2n+1)^{2}\pi ^{2}\nu t}
{4d^{2}}\right],
\end{eqnarray}
where $d$ is half the distance between the parallel plates, $\nu =\eta /\rho$ is
the kinematic viscosity, and $F=-2\nu v_{0}/d^{2}$ is a driving force proportional
to the pressure difference ($\Delta p$) between the inlet and outlet, and
$v_{0}=-d^{2}\Delta p/(2\rho\nu L)$ is a constant asymptotic velocity, where
$L$ is the length of the pipe section. In the limit when $t\to\infty$ the above
solution tends to the well-known steady-state, parabolic profile 
\begin{equation}
v_{x}(y)=v_{0}\left(1-\frac{y^{2}}{d^{2}}\right).
\end{equation}
For this test problem we choose $\rho _{0}=1000$ kg m$^{-3}$,
$v_{0}=1.25\times 10^{-5}$ m s$^{-1}$, and $\nu =1.0\times 10^{-6}$ m$^{2}$ s$^{-1}$,
corresponding to a Reynolds number Re$=2dv_{0}/\nu =0.0125$.
The fluid domain is filled with 1942 particles initially at rest and regularly
distributed in the spanwise direction between $x=0$ and $x=L=2.33\times 10^{-4}$ m.
The particles are given a smoothing length $h\approx 2.4\times 10^{-5}$ m and Eq. (4)
is used as the pressure-density relation with $c_{0}=2$ m s$^{-1}$. 
\begin{figure}
\centering
\includegraphics[width=10cm]{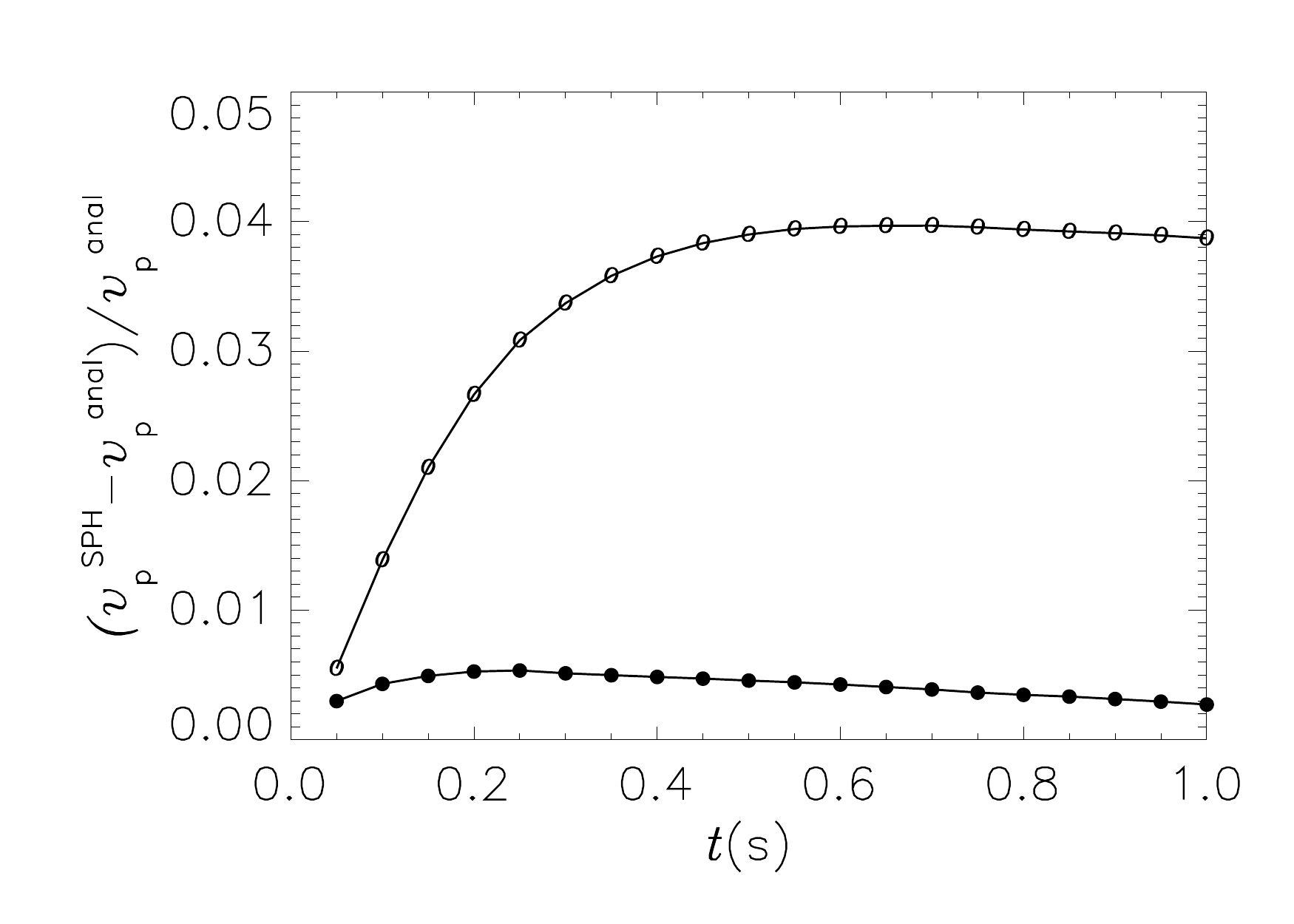}
\caption{Relative error of the SPH peak velocity, $v_{\rm p}^{\rm SPH}$, as compared
with the theoretically predicted value, $v_{\rm p}^{\rm anal}$, as a function of time.
The dots correspond to the calculation of Fig. 2 with nonreflecting outlet boundary
conditions. For comparison, the circles represent the resulting relative errors when
periodic boundary conditions are used for the same test.}
\end{figure}
\begin{figure*}
\centering
\includegraphics[width=10cm]{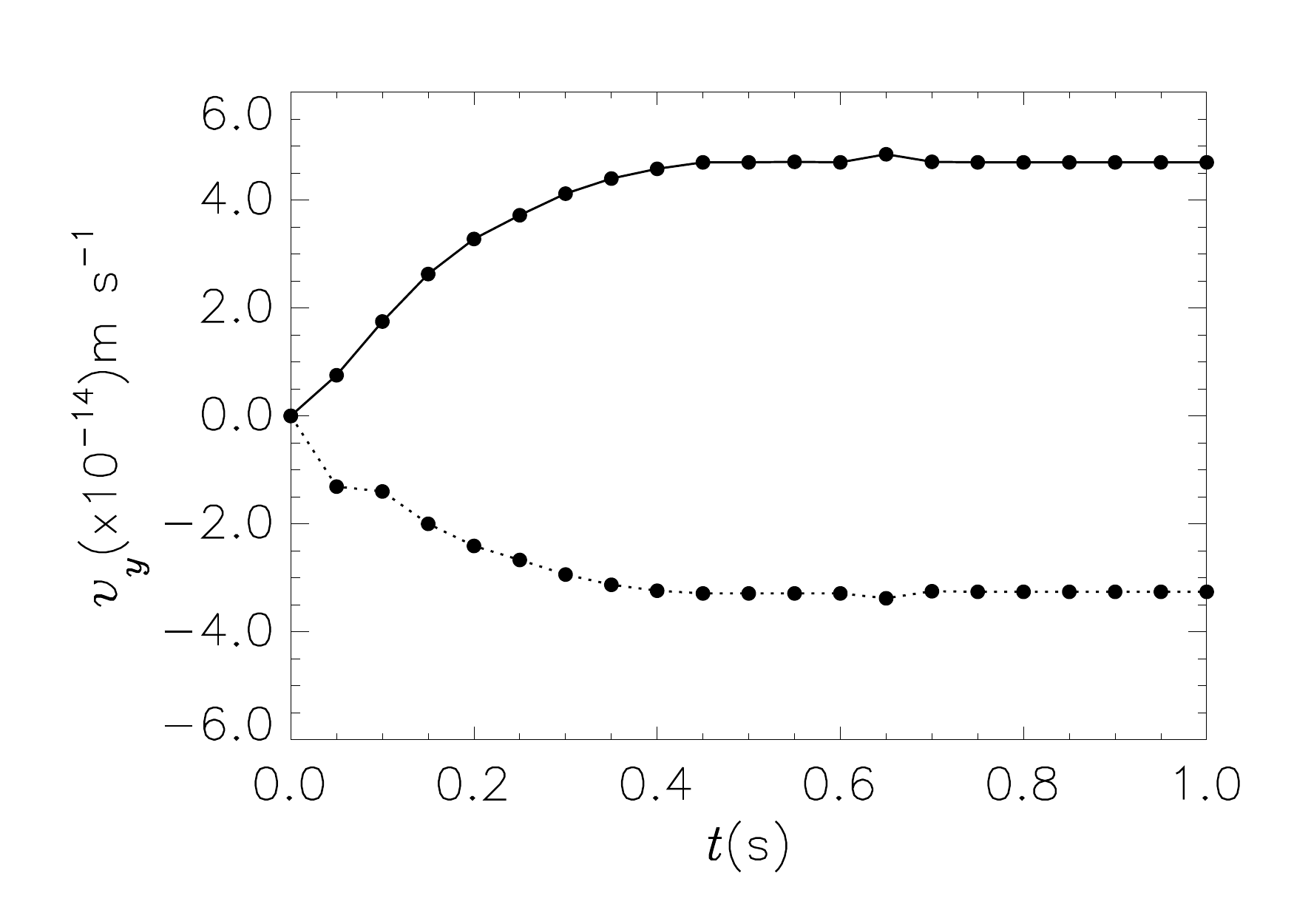}
\includegraphics[width=10cm]{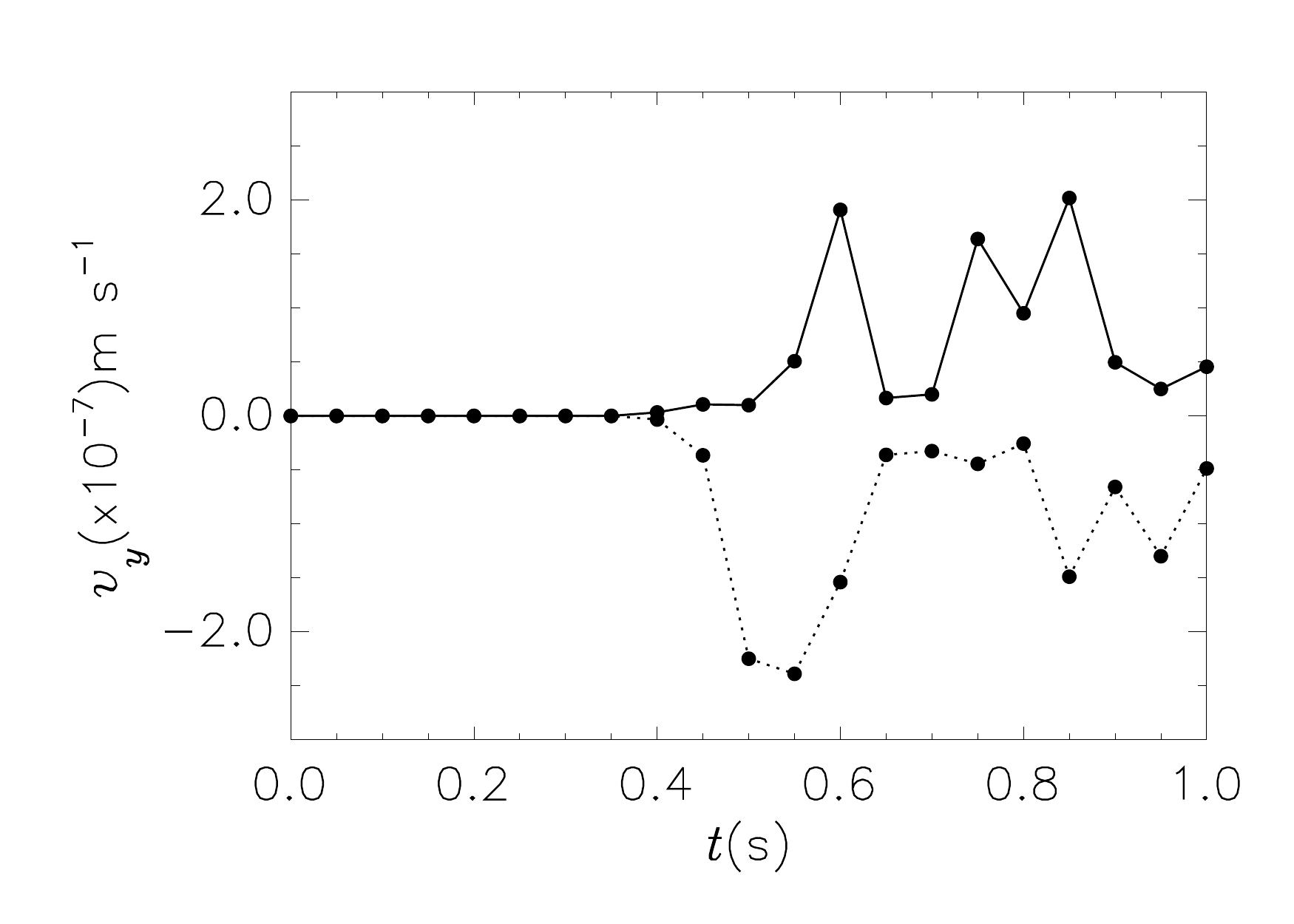}
\caption{Maximum values of the numerically induced $y$-component of the velocity
above (dotted-solid line) and below (dotted-dashed line) the central plane ($y=0$)
as a function of time for the plane Poiseuille flow of Fig. 2 using nonreflecting
outlet (top frame) and periodic boundary conditions (bottom frame). The level of
anisotropy is about seven orders of magnitude lower for the nonreflecting outlet
simulation.}
\end{figure*}

The transient behaviour as calculated with the outlet boundary condition treatment
is shown in Fig. 2, where the SPH solution (dots) is compared with the theoretically
predicted one (solid lines) at selected times. The solution is depicted up to $t=1.0$ s
when a steady-state regime has been already established. Figure 3 shows the relative
errors between the numerical and analytical peak velocities for the run of Fig. 2
with the outflow boundary condition method (dots) as compared with an identical run using
periodic boundary conditions (circles). The error in the periodic simulation grows
rapidly during the first 0.4 s and reaches values that are an order of magnitude 
greater than the error carried by the nonreflecting outlet simulation.
This difference is the result of cumulative errors due to 
the recycling of numerical disturbances in the periodic simulation. Figure 4 also
shows the numerical $y$-component of the velocity for both calculations. With the
present method (top frame), the numerically induced $y$-component of the velocity
reaches maximum absolute values above and below $y=0$ of $\approx 4.0\times 10^{-14}$ 
m s$^{-1}$, while in the periodic case (bottom frame)
$v_{y}$ exhibits erratic oscillations after $\sim 0.4$ s and reaches values
that are about 7 orders of magnitude higher. The asymmetry of the oscillations with
respect to the central plane $y=0$ is indicative of the presence of noise due to
the periodic recycling of particles. Evidently, the nonreflecting outlet boundary
conditions are doing a superior job for this test as the flow remains laminar with
a very good accuracy.

\subsection{Flow between inclined plates}

\begin{figure}
\centering
\includegraphics[width=12cm]{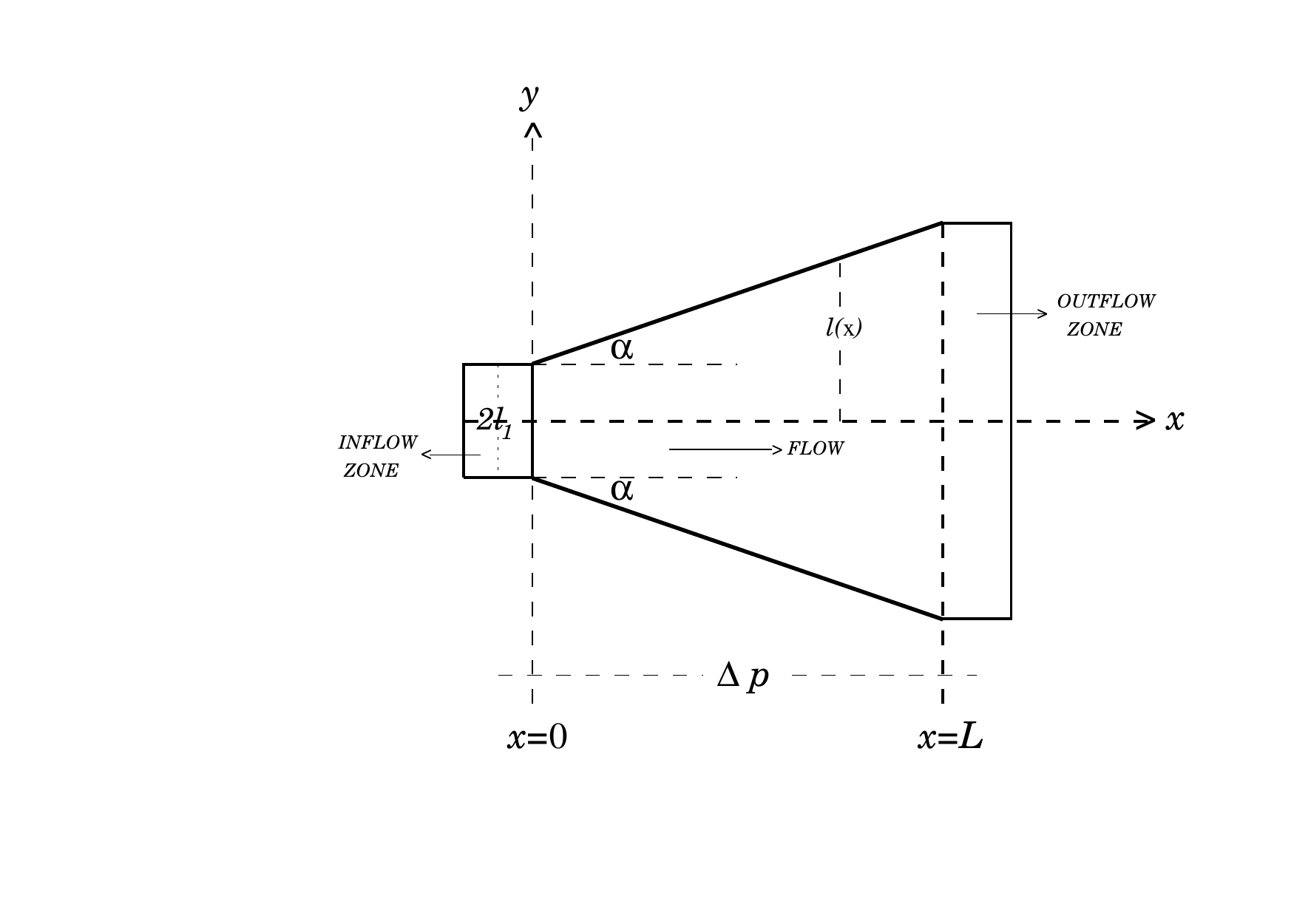}
\caption{Top view of the flow geometry between two inclined plates.}
\end{figure}

As a second test, we consider the case of laminar flow between two inclined plates,
where the inlet and outlet cross-sections differ as shown in Fig. 5. If the
inclination angle $\alpha$ is small and the flow is driven by a pressure difference
between the inlet and outlet planes, an analytical solution can be derived for the
streamwise velocity after a steady-state flow is reached \cite{Lin05,Liang12},
namely
\begin{equation}
v_{x}=-\frac{\Delta p}{\eta L}\left[y^{2}-\left(l_{1}+x\tan\alpha\right)^{2}\right]
\frac{l_{1}^{2}\left(l_{1}+L\tan\alpha\right)^{2}}{\left(2l_{1}+L\tan\alpha\right)
\left(l_{1}+x\tan\alpha\right)^{3}},
\end{equation}
where $\eta =\rho\nu$ is the shear viscosity, $l_{1}$ is half the separation of the
inclined plates at the inlet, and $L$ is the distance between the inlet and outlet
planes. Following the procedure described by Liang et al. \cite{Liang12}, the body
force at the inlet $x=0$ is given by
\begin{equation}
F(x=0)=\frac{2\Delta p\tan\alpha}{\rho l_{1}^{3}}\left[\frac{1}
{\left(l_{1}+L\tan\alpha\right)^{2}}-\frac{1}{l_{1}^{2}}\right]^{-1},
\end{equation}
while at any streamwise position $x$ it obeys the relation
\begin{equation}
F(x)=\left[\frac{l_{1}}{l(x)}\right]^{3}F(x=0),
\end{equation}
where $l(x)=l_{1}+x\tan\alpha$. This gives $F(x=L)=F(x=0)/(1+L\tan\alpha /l_{1})^{3}$
at the outlet. Equations (33)--(35) are valid if the Reynolds number Re$<1$ and
$L\gg l_{1}$.
\begin{figure}
\centering
\includegraphics[width=10cm]{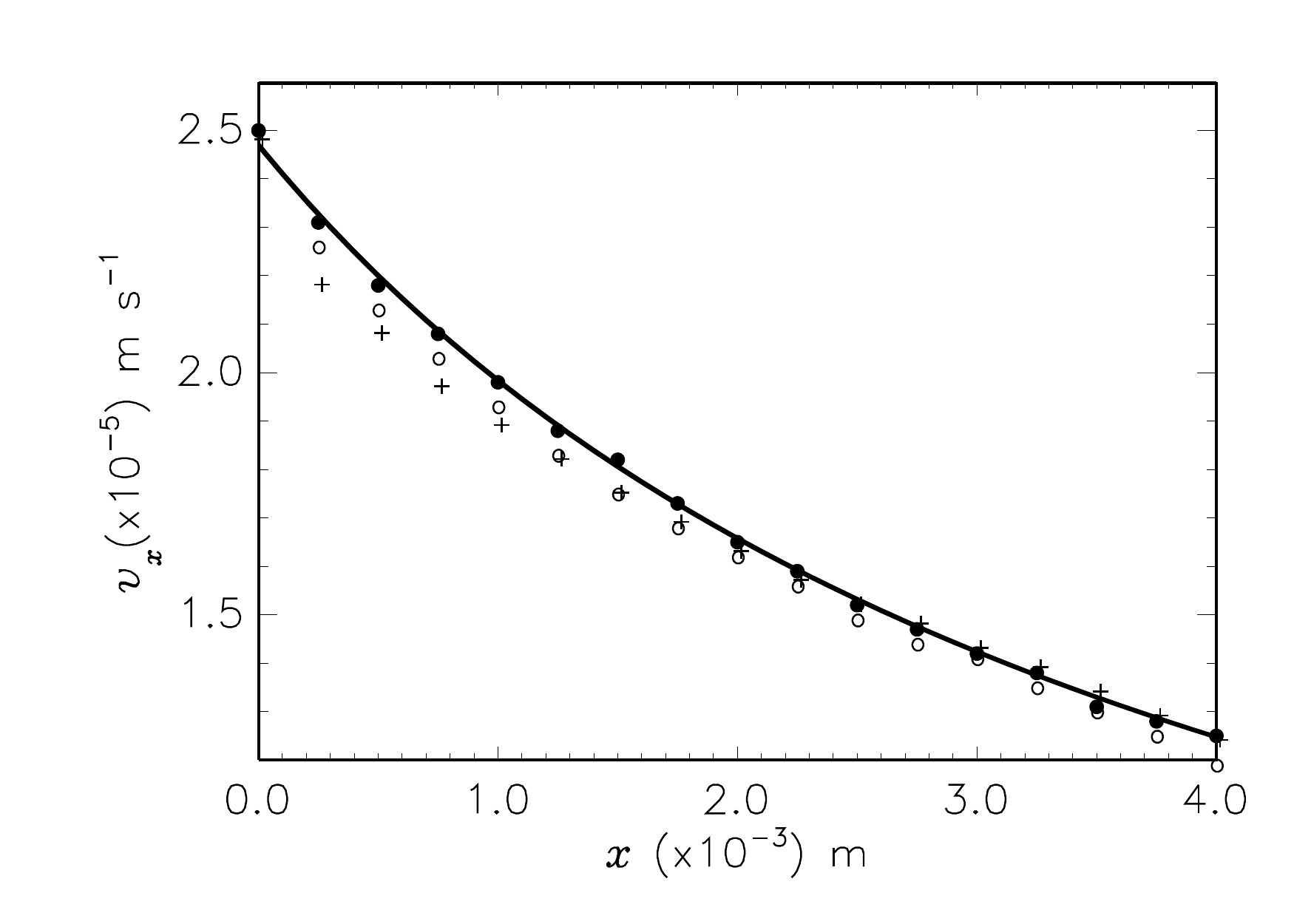}
\caption{Steady-state streamwise velocity profile as a function of position along
the inclined plates. The analytical solution given by Eq. (33) (solid line) is
compared with the numerically obtained profiles at different spatial resolutions:
$N=6592$ (crosses), 15984 (circles), and 103168 (dots).}
\end{figure}

For this test case we choose the same parameters as in Liang et al. \cite{Liang12},
that is, $\Delta p\approx -1.217\times 10^{-3}$ N m$^{-2}$, $L=4$ mm, $2l_{1}=0.5$ mm,
$\nu =1.0\times 10^{-6}$ m$^{2}$ s$^{-1}$, $\rho =1000$ kg m$^{-3}$, and 
$\alpha =3.503^{\circ}$, except for the sound speed which is taken to be
$c=5.0$ m s$^{-1}$ in order to keep the density fluctuations below 1\% with the use
of Eq. (4). In contrast, Liang et al. \cite{Liang12} used
an equation of state of the form $p=c^{2}p$ with $c=2.5\times 10^{-4}$ m
s$^{-1}$. Initially the particles are at rest and distributed on a regular Cartesian
mesh and the initial smoothing length is set to $h=1.1\Delta$, where $\Delta$ is the
initial interparticle distance along the $x$- and $y$-directions. With the above
parameters, the Reynolds number of the flow is Re$=2l_{1}v_{0}/\nu =0.0125$, where
$v_{0}=2.5\times 10^{-5}$ m s$^{-1}$ is the velocity at the inlet plane. According
to Eq. (34), the body force at the inlet is $F(x=0)=8.0\times 10^{-4}$ m s$^{-2}$,
while the body force entering in Eq. (2) is ${\bf F}={\bf x}F(x)$, which is always
parallel to the $x$-axis and zero otherwise.

\begin{figure}
\centering
\includegraphics[width=12cm]{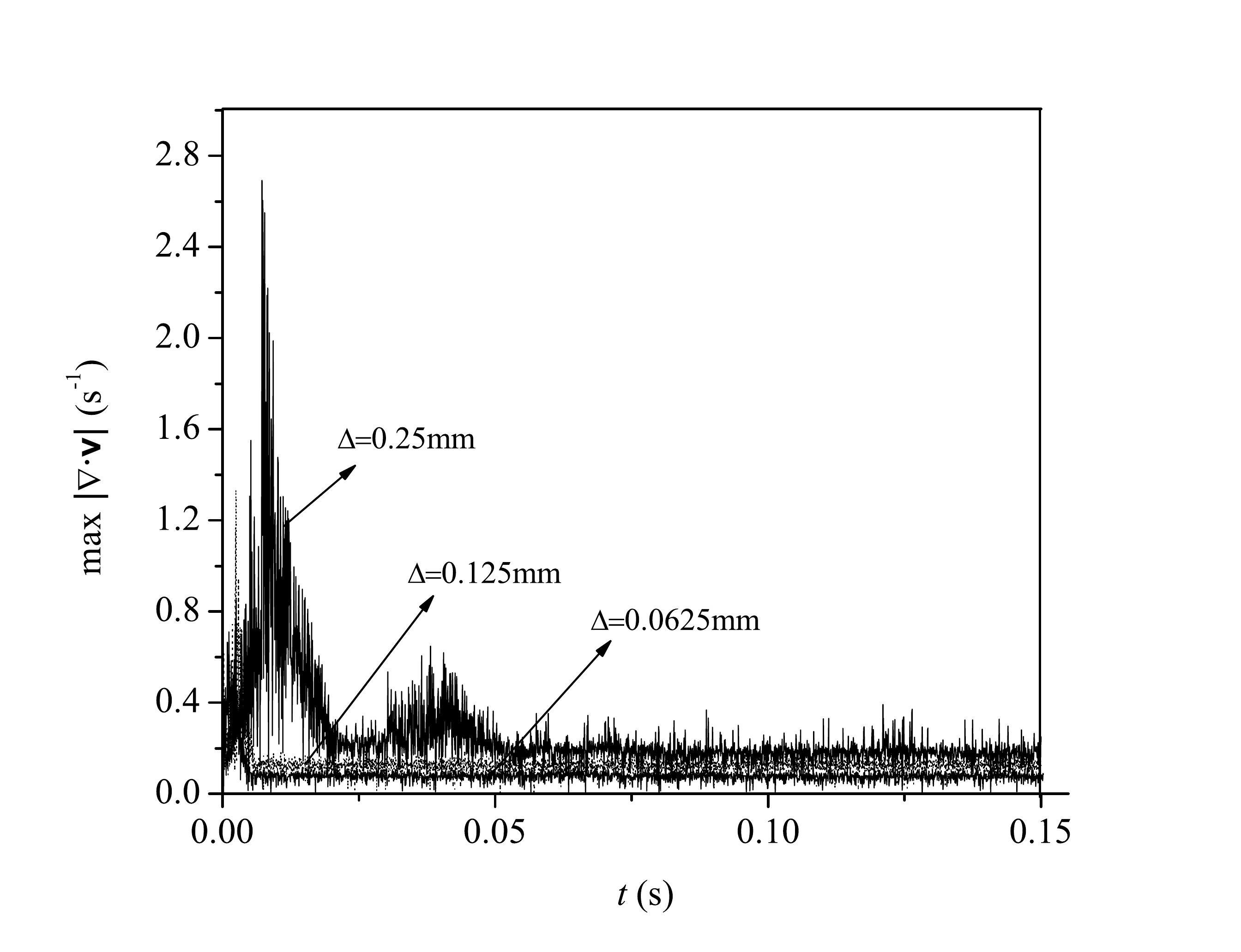}
\caption{Absolute value of the maximum divergence of the velocity field as a function
of time for flow between two inclined plates. The results for the three resolutions
tried are shown.}
\end{figure}

Figure 6 depicts the $x$-component of the fluid velocity after about 0.15 s when the
flow reaches a steady state. The numerically obtained profiles are compared with the
analytical solution (solid line) as given by Eq. (33). Three different runs are
shown with identical initial parameters but varied initial spatial resolution:
$N=6592$ (crosses), 15984 (circles), and 103168 (dots), corresponding to initial
interparticle separations of 0.25, 0.125, and 0.0625 mm, respectively. Figure 6 shows
that the results obtained with the present method converge to the theoretical
solution as the resolution is increased. In terms of the root-mean-square error
(RMSE)
\begin{equation}
{\rm RMSE}(v_{x})=\sqrt{\frac{1}{N}\sum _{a=1}^{N}\left(v_{x,a}^{\rm anal}-
v_{x,a}^{\rm SPH}\right)^{2}},
\end{equation}
where $v_{x,a}^{\rm anal}$ represents the analytical solution (33) at the position
of particle $a$ and $v_{x,a}^{\rm SPH}$ the corresponding SPH calculated value, the
numerical errors decrease with decreasing initial particle size with 
RMSE$(v_{x})\approx 1.49\times 10^{-6}$ m s$^{-1}$ (for $\Delta =0.25$ mm), 
$\approx 1.06\times 10^{-6}$ m s$^{-1}$ (for $\Delta =0.125$ mm), and
$\approx 1.54\times 10^{-7}$ m s$^{-1}$ (for $\Delta =0.0625$ mm).

The pressure constant $p_{0}$ in Eq. (4) governs the relative density fluctuations
$|\Delta\rho|/\rho _{0}$, with $\Delta\rho =\rho -\rho _{0}$. Since
$|\Delta\rho|/\rho _{0}\sim M^{2}$, where $M$ is the Mach number, density fluctuations
in the flow can be kept of the order of 1\%, or less, by choosing $M\leq 0.1$. To enforce
this condition $p_{0}$ must be equal to $c_{0}^{2}\rho _{0}/\gamma$, where $c_{0}$ is the
sound speed at the reference density $\rho _{0}$ which is chosen large enough to guarantee
that $M\leq 0.1$. Figure 7 shows the maximum value of $|\nabla\cdot {\bf v}|$ in the flow
as a function of time for the three resolutions tried. During the first 0.05 s, peaks of
the velocity divergence as high as $\sim 2.7$ s$^{-1}$ and $\sim 0.6$ s$^{-1}$ arise
in the low resolution run. At later times, the maximum velocity divergence decreases
and oscillates about 0.2 s$^{-1}$. As the resolution is increased to $\Delta =0.1250$ mm,
the peak intensity at the beginning is reduced to less than $\sim 0.8$ s$^{-1}$ and the
maximum value of the divergence oscillates about $\sim 0.12$ s$^{-1}$. This mean value
improves to $\sim 0.08$ s$^{-1}$ for the high resolution run. In this case, the divergence
achieves a peak of $\sim 0.4$ s$^{-1}$ at the very beginning. The actual maximum density
fluctuations associated with these deviations from exact incompressibility (calculated as
the product $\max(|\nabla\cdot {\bf v}|_{a})\Delta t$) correspond to mean values of
$3.2\times 10^{-7}$ (for $\Delta =0.25$ mm), $2.6\times 10^{-7}$ (for $\Delta =0.125$ mm),
and $9.1\times 10^{-8}$ (for $\Delta =0.0625$ mm).

\subsection{Kelvin-Helmholtz instability in a channel}

We now assess the ability of our method to inhibit feedback noises when convecting 
flow anisotropies across the outlet. The test case concerns the onset of the 
Kelvin-Helmholtz (KH) instability at the interface between two shearing fluids of
different velocities when velocity perturbations perpendicular to the interface grow
to eventually mix the layers \cite{Chandrasekhar81}. We consider a two-dimensional
setup similar to that reported by Price \cite{Price08}, using $N=10880$ equal mass
particles filling the domain $0\leq x\leq 0.4$ m and $0\leq y\leq 0.1442$ m. The
particles are initially placed on a uniform Cartesian array and the density is set
to $\rho =1000$ kg m$^{-3}$ everywhere.
\begin{figure}
\centering
\includegraphics[width=11cm]{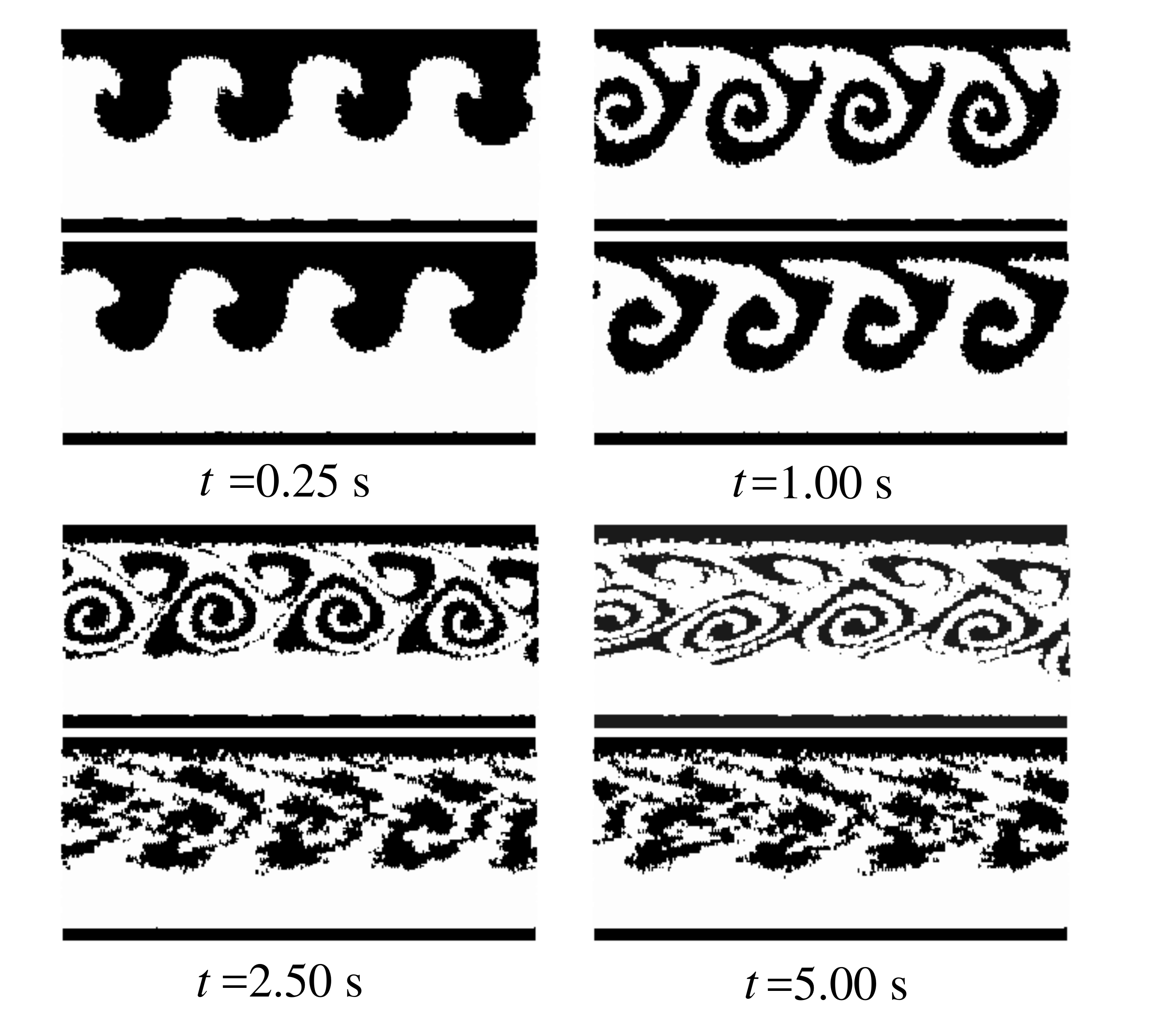}
\caption{Density maps of the two-dimensional KH instability at different times.
At each time, the top and bottom frames correspond to identical simulations
using the nonreflecting outlet and periodic boundary conditions in the 
$x$-direction, respectively. Except for small features, the linear and early 
non-linear growth is similar in both simulations up to about 1.0 s.}
\end{figure}

A shear flow is setup in the $x$-direction with velocity $v_{x}=1$ m s$^{-1}$ for
$0\leq y<0.103$ m and $v_{x}=2$ m s$^{-1}$ for $0.103\leq y\leq 0.1442$ m, so
that the tangential fluid velocity has a discontinuous jump across the interface
between the streams. This flow corresponds to ${\rm Re}=10000$. For this test
we use Eq. (4) with $c_{0}=40$ m s$^{-1}$. This configuration is known to be
susceptible to a KH instability at all wavelengths. The instability is seeded by 
introducing a small velocity in the $y$-direction given by
\begin{equation}
v_{y}=A\sin\left[\frac{-2\pi}{\lambda}\left(x+\frac{1}{2}\right)\right],
\end{equation}
for $0.09<y<0.116$ m and zero elsewhere, with $A=0.5$ m s$^{-1}$ and $\lambda =0.1$ m.
For this setup the linear KH growth time-scale for the sinusoidal mode defined by
\begin{equation}
\tau _{\rm KH}=\frac{2\lambda}{|v_{x,1}-v_{x,2}|},
\end{equation}
is $\tau _{\rm KH}=0.2$ s. No-slip boundary conditions are applied at contact with
the walls of the channel. For this test the inlet consists of an upstream
section of 0.8 m long, while the actual channel has a length of 0.4 m and the outflow
zone is 0.1 m long. Initially, the inlet and the channel sections are filled with
particles, which are then evolved from the above initial conditions using SPH. 
As the flow proceeds, the inlet section becomes progressively depleted of particles,
resembling a moving piston boundary condition. The calculation is halted immediately
before the inlet becomes completely depleted of particles. At the exit of the channel,
the outlet boundary conditions are employed. For
this test calculation, Eq. (5) is used with the viscous force term replaced by an
artificial viscosity using the scheme proposed by Monaghan \cite{Monaghan92} with a
coefficient $\alpha _{\nu}=0.01$. In order to test the performance of the nonreflecting 
outlet boundary conditions, a second run using periodic boundary conditions at the
inlet and outlet in the $x$-direction was performed for direct comparison.
\begin{figure}
\centering
\includegraphics[width=10cm]{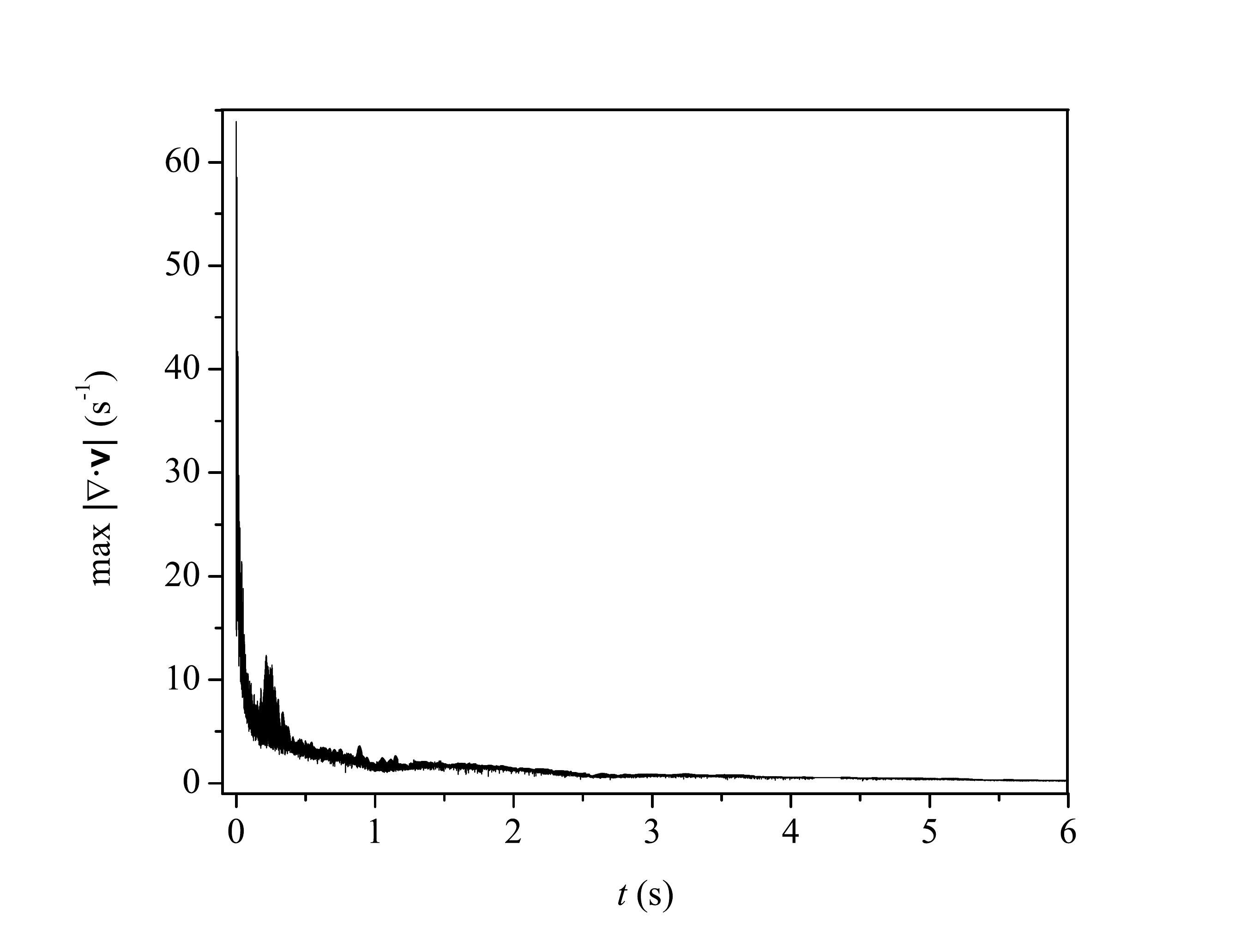}
\caption{Absolute value of the maximum velocity divergence as a function of time for
the KH instability simulation of Fig. 8 using nonreflecting outlet boundary conditions.}
\end{figure}
\begin{figure}
\centering
\includegraphics[width=10cm]{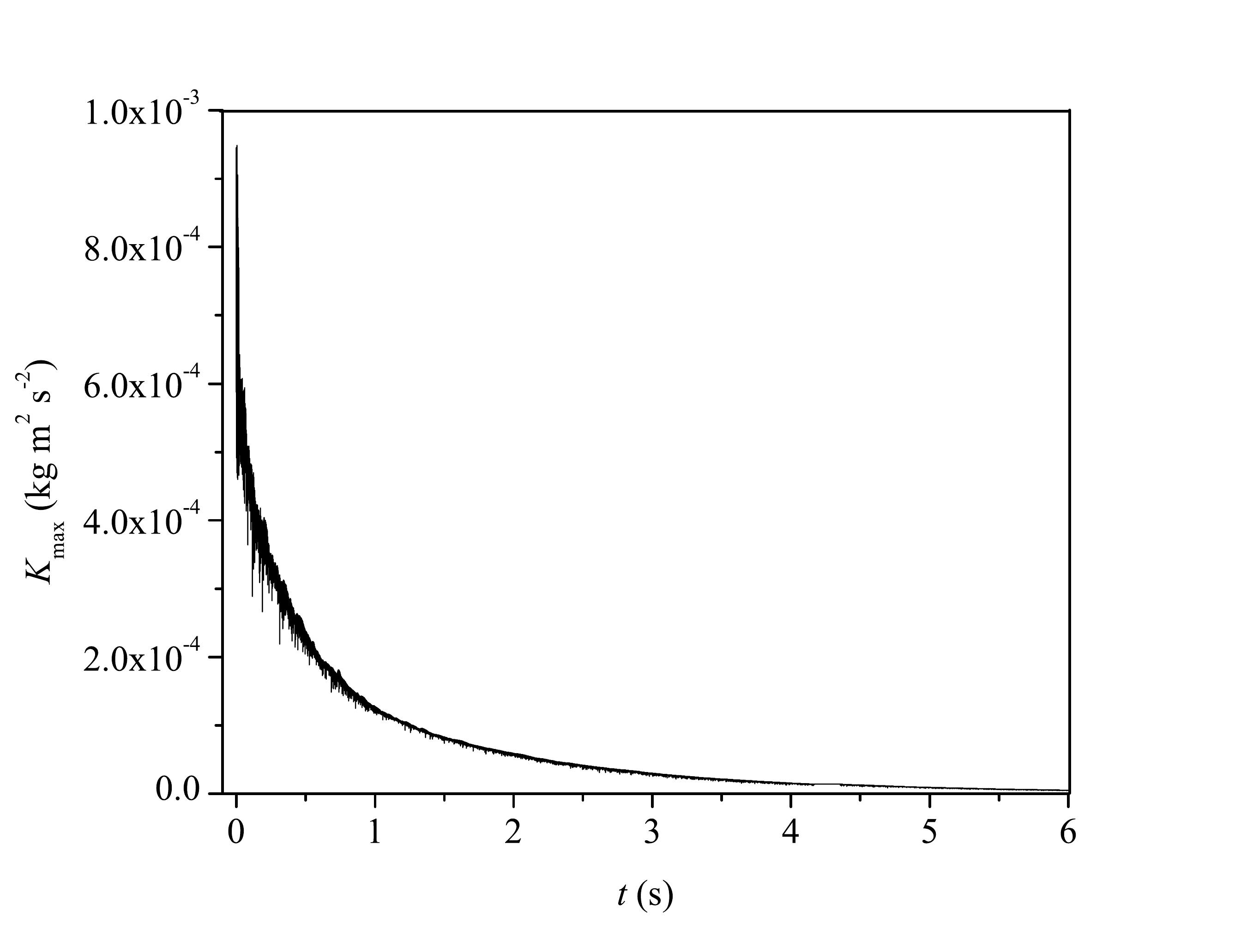}
\caption{Maximum value of the kinetic energy as a function of time for the KH 
instability simulation of Fig. 8 using nonreflecting outlet boundary conditions.}
\end{figure}
Figure 8 shows the results at different times up to 5.0 s. At each time, the top
and bottom frames correspond to nonreflecting outlet and periodic simulations,
respectively. In the former case 5508 inflow particles were needed to follow the
evolution up to 5.0 s by which time the inflow zone was almost depleted.
With the nonreflecting boundary conditions the linear growth phase is similar
to the periodic simulation. The instability grows at the shear layer and
the peaks of each fluid phase penetrate into each other ($t=0.25$ s). After further
penetration of the fluid phases, non-linear shear leads them to roll up
into the well-known KH whorls ($t=1.0$ s). We may see that the whorl height is
nearly identical in both caculations. However, at $t=1.0$ s the rolling appears
to be slightly more pronounced in the nonreflecting outlet simulation. At later
times, the interface rolls up into a sequence of spiral vortices ($t=2.5$ s). As
time progresses, the turns are elliptically deformed ($t=5.0$ s). While a vortex
field is formed with the nonreflecting boundary conditions, which then amplifies
and eventually leads to mixing, the solution with periodic boundary conditions 
looks highly degraded by $t=2.5$ s because of the continued re-entry of numerical
perturbations.

Figure 9 depicts the time evolution of the absolute value of the maximum
velocity divergence for the nonreflecting outlet simulation. At the very beginning
the velocity divergence drops sharply, decaying from $\sim 67$ s$^{-1}$ to less
than $\sim 0.3$ s$^{-1}$ during the first second of the evolution. After this
time, it decreases slowly to less than $\sim 0.2$ s$^{-1}$ by $t=6.0$ s.
In addition, Fig. 10 shows the time evolution of the maximum kinetic energy.
During the first two seconds, the maximum kinetic energy is seen to decrease 
rapidly by an order of magnitude and then at a much slower rate during the
spiraling and elliptical deformation of the vortex sheet, reaching a value of
$\sim 4.0\times 10^{-6}$ kg m$^{2}$ s$^{-2}$ by 6.0 s, when the calculation is
terminated because of particle depletion in the inlet section.  

\subsection{Flow through a constricted channel}

\begin{figure}
\centering
\includegraphics[width=10cm]{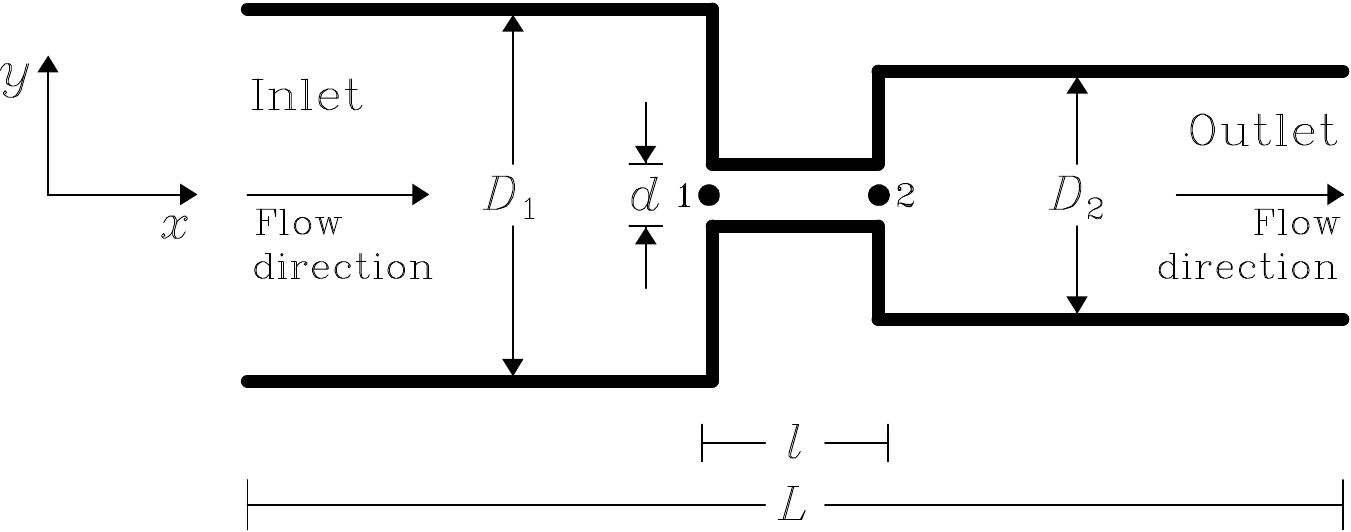}
\caption{Top view of the constricted channel section for choked flow.}
\end{figure}

As a further test we consider the flow between two parallel walls with a sharp-edged,
narrow passage (or throat) of length $l$ at the centre of the channel, as shown in
the top view of Fig. 11. The main flow direction is taken along the $x$-axis and
the depth of the channel is assumed to be infinite so that the flow is in the
($x$,$y$)-plane. Three separate simulations are considered. Two of them use identical
parameters except that in one run (R1) particles in the outflow zone are evolved
solving Eq. (29) with the anisotropic term dropped to mimic an Orlanski type
outlet boundary condition, while the second run (R2) solves Eq. (29) including the
anisotropic term. A third run (R3) is identical to R2 but with a longer 
downstream pipe section. This test problem is more
stringent than the previous examples because downstream the throat anisotropic
flow develops as in the case of flow past a backward facing step. In addition, if
the throat is modelled as a very narrow passage, its cross-section can be made to
strongly differ from that of the outlet as desired.
\begin{figure}
\centering
\includegraphics[width=15cm]{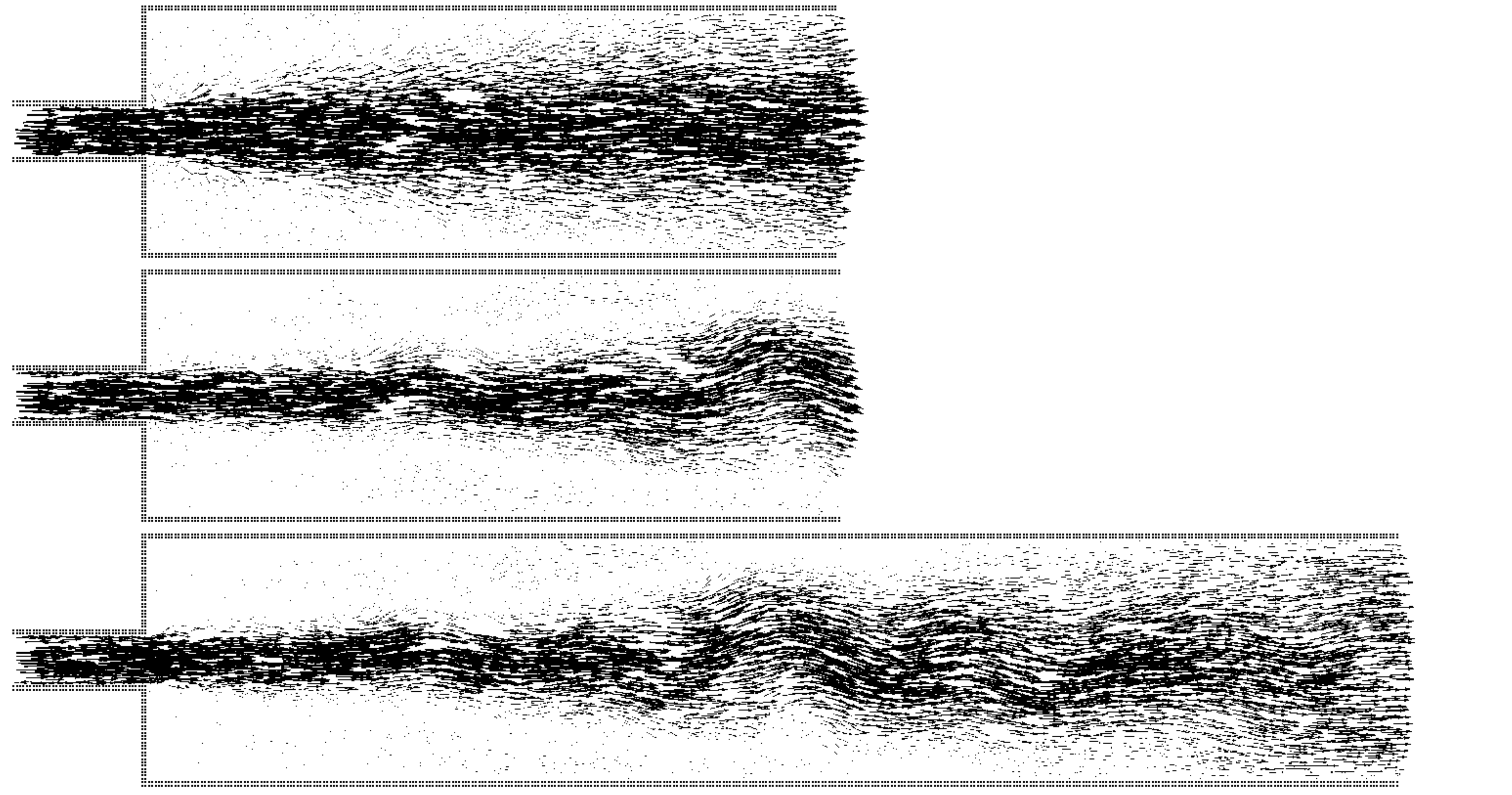}
\caption{Velocity vectors in the throat and downstream sections of the channel at
$t=4.65$ s for three separate simulations: the R1 run using an Orlanski type outlet
boundary condition with no anisotropic term in Eq. (29) (top frame), the
R2 run with the anisotropic term in Eq. (29) included (middle frame), and the R3 run, 
which is identical to R2 but with a longer downstream section (bottom frame). A
very good agreement for R2 and R3 is shown, proving the efficiency of the 
nonreflecting type boundary conditions implemented by solving the outgoing wave
Eq. (29).}
\end{figure}

For these simulations we take $\rho =1000$ kg m$^{-3}$, $\nu =1.0\times 10^{-6}$
m$^{2}$ s$^{-1}$, and a time-varying plane Poiseuille velocity profile in the inflow
zone given by
\begin{equation}
v_{\rm in}=v_{0}\left(\frac{t}{t_{0}}\right)\left(1-\frac{4y^{2}}{D_{1}^{2}}\right),
\end{equation}
where $t_{0}=0.2$ s and $v_{0}=4.45\times 10^{-2}$ m s$^{-1}$. The inlet flow is
exactly zero at $t=0$ and increases linearly with time in the course of the evolution.
This is equivalent to applying a pressure difference by suction on the outlet. When
$t\geq t_{0}$, we set $t/t_{0}=1$ and the inlet flow becomes time-independent. The
sound speed is taken to be $c_{0}=1.0$ m s$^{-1}$ and Eq. (4) is used as the
pressure-density relation. The throat has a length of $l=7.78$ cm and an opening
width of $d\approx 1.33$ cm. The upstream section has a width of $D_{1}\approx 8.89$
cm and a length $l_{1}=20$ cm, while the downstream section has a width of
$D_{2}\approx 7.56$ cm and a length $l_{2}=20$ cm for models R1 and R2 and $l_{2}=36$ cm
for model R3. For these simulations we use a total of 37467 (for models R1 and R2) and
50346 regularly distributed particles (for model R3) filling the entire channel.
\begin{figure}
\centering
\includegraphics[width=15cm]{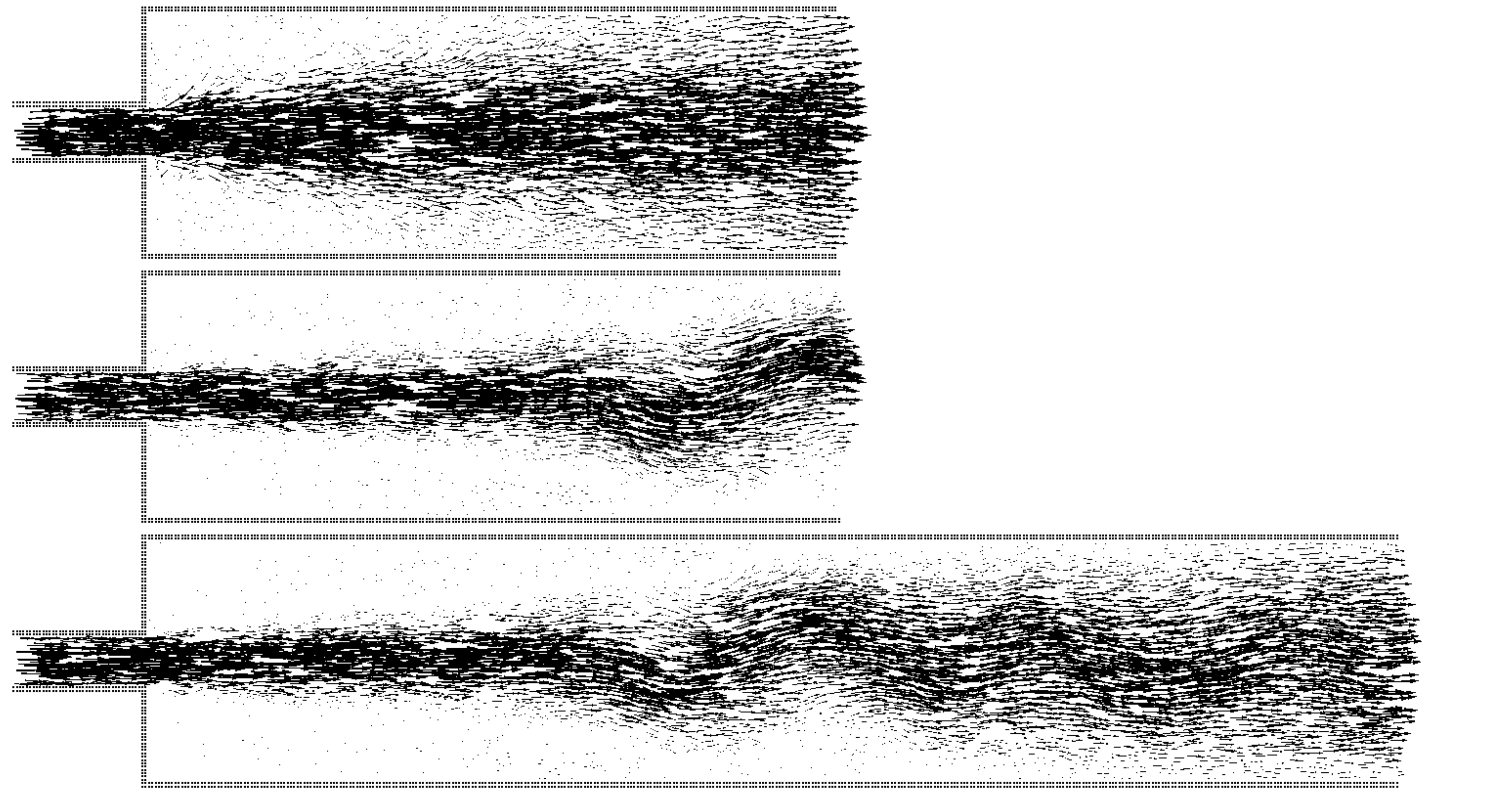}
\caption{The same as Fig. 12 but at $t=6.0$ s.}
\end{figure}
\begin{figure}
\centering
\includegraphics[width=14cm]{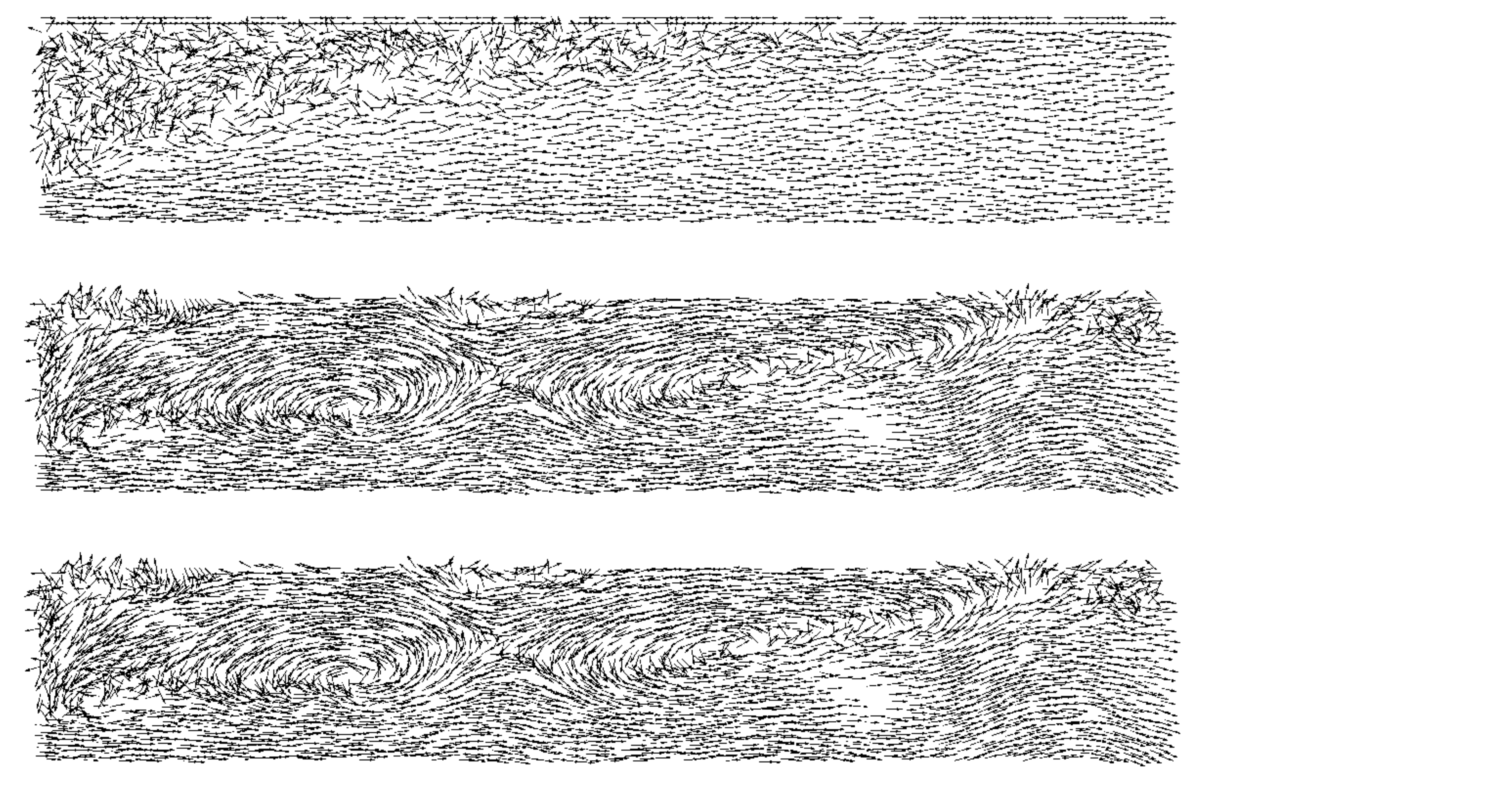}
\caption{Blowup view of the downstream section showing details of the flow on the
side just above the vena contracta at $t=4.65$ s for the three models of Fig. 12.
To facilitate direct comparison, the lenght of the downstream section used for models 
R1 (top frame) and R2 (middle frame) is also shown for model R3 (bottom frame).} 
\end{figure}

The influence of the computational domain size and type of nonreflecting boundary
conditions as given by Eqs. (27) and (28) are now examined. Figures 12 and 13 display
the velocity field in the throat and downstream sections at $t=4.65$ and 6.0 s,
respectively. The top frame shows the flow structure for model R1 with the Orlanski
type outlet boundary, while the other two frames correspond to models R2 (middle
frame) and R3 (bottom frame) using nonreflecting conditions with the anisotropic
term in Eq. (29) included and a different size of the downstream section. The Reynolds
number in the throat conduit can be defined as Re$=v_{m}d/\nu$, where $v_{m}$
is the mean velocity there. This gives Re$\approx 4531$ (at $t=4.65$ s) and
$\approx 4552$ (at $t=6.0$ s) for cases R2 and R3, while Re$\approx 4516$ (at
$t=4.65$ s) and Re$\approx 4546$ (at $t=6.0$ s) for model R1. At $t=6.0$ s, the maximum
velocity at the exit of the throat is $v_{max}\approx 0.43$ m s$^{-1}$ for model R1 against
$v_{max}\approx 0.40$ m s$^{-1}$ for models R2 and R3, while the mean pressure drop
through the throat is $\Delta p\approx 6.51\times 10^{-3}$ Pa for model R1 compared to
$\Delta p\approx 1.22\times 10^{-2}$ Pa for the other two cases. 
In the downstream section, a jet forms just behind the throat exit surrounded by
recirculatory flow, which extends along the full length of the section. Winding
of the jet downstream is due to its interaction with the moving smallest vortices.
Details of this recirculatory flow are displayed in Fig. 14, which show blowup
views of the flow just above the vena contracta at $t=4.65$ s for the models of Fig.
12. Inspection of these figures shows that the vortices appearing downstream on both sides
of the vena contracta are damped in model R1 (top frame) compared to models R2 (middle
frame) and R3 (bottom frame), implying that neglecting the anisotropic term in Eq. (29)
affects the structure of the flow. Also the reattachment length at $t=4.65$ and 6.0 s 
is much shorter in model R1 ($\approx 0.10$ m) compared to model R3 ($\approx 0.24$ m).
As shown in Figs. 12 and 13, the flow structure for model R2 closely follows that
shown for model R3 with a longer downstream channel, demonstrating that the feedback
noise from the outlet boundary is also greatly reduced for this test.
\begin{figure}
\centering
\includegraphics[width=10cm]{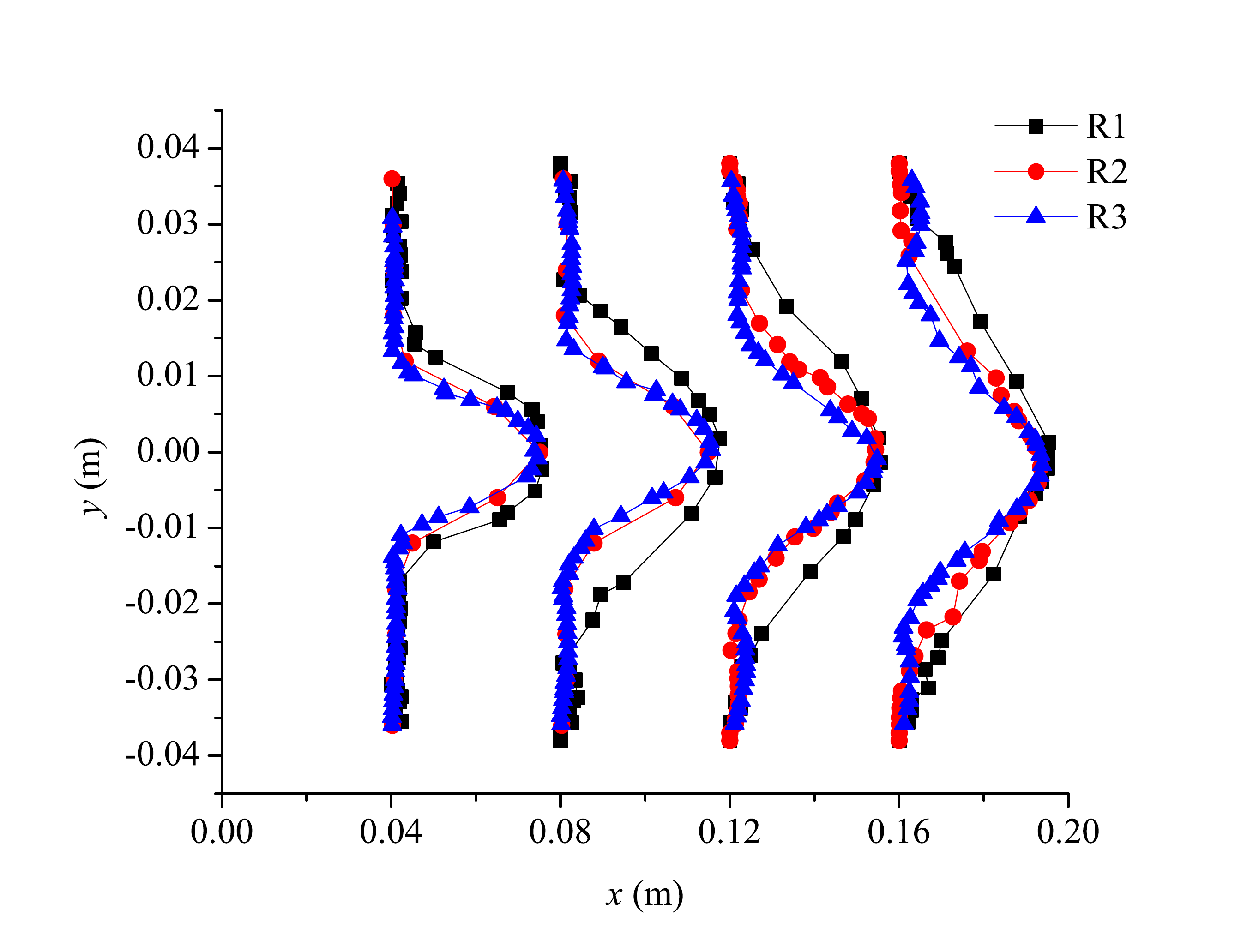}
\caption{Comparison of the streamwise velocity profiles at successive stations in
the downstream section of the constricted channel for models R1, R2, and R3 at
$t=4.65$ s.}
\end{figure}
\begin{figure}
\centering
\includegraphics[width=10cm]{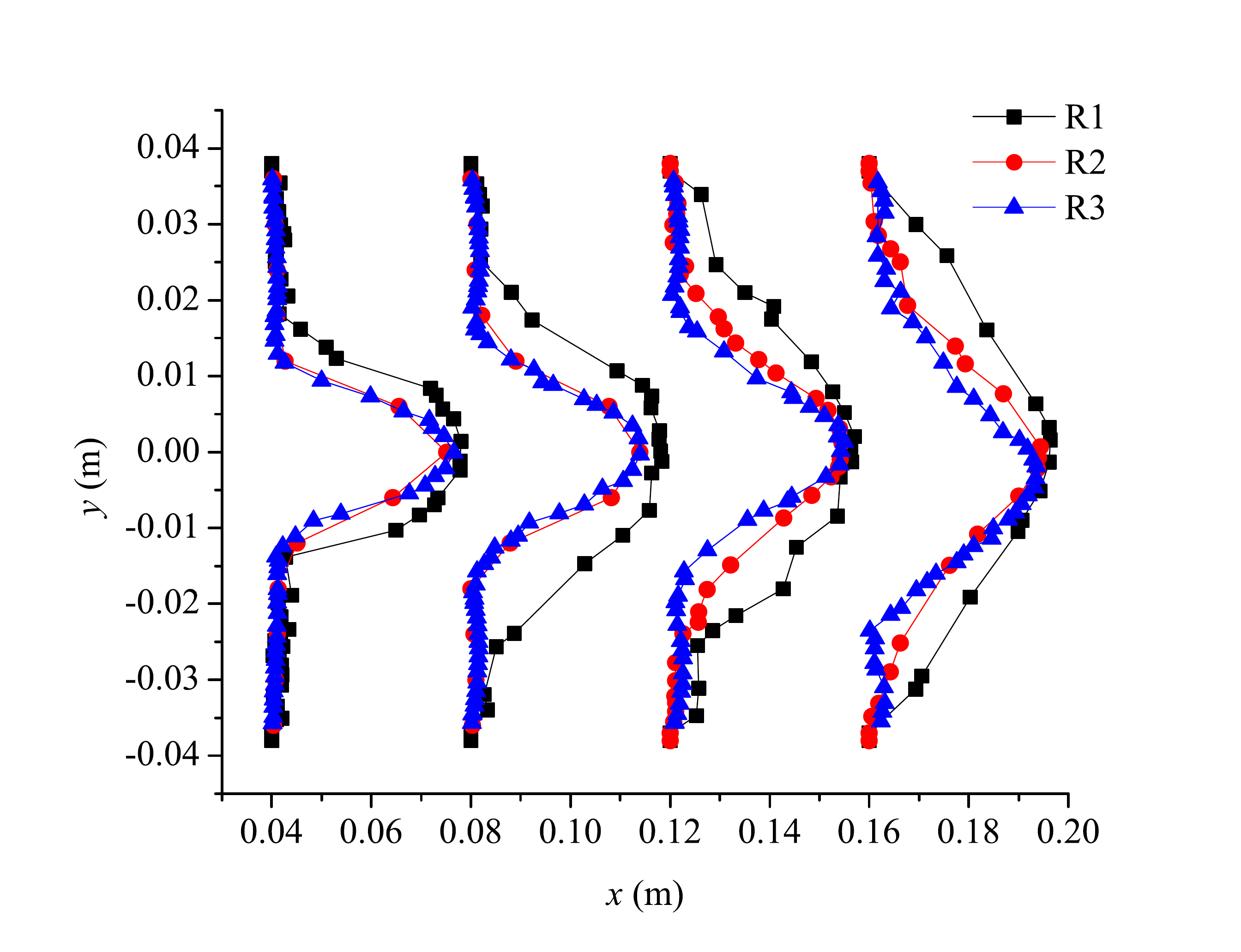}
\caption{The same as Fig. 15 at $t=6.0$ s.}
\end{figure}

Figures 15 and 16 show longitudinal velocity profiles for models R1, R2, and
R3 at successive streamwise stations along the downstream section for the same
times of Figs. 12 and 13, respectively. In both figures, the squares depict the
profiles for model R1, while the dots and triangles correspond to the profiles for
models R2 and R3, respectively. It can be clearly seen that there is a very good
correspondence between the profiles for models R2 and R3 for all stations at both
times. In contrast, the profiles of model R1 substantially deviate from those
of models R2 and R3 on both sides of the centreline and towards the channel walls, 
with the magnitude of the deviations
increasing close to the outlet. The good correspondence between the results of
models R2 and R3 proves the efficiency of the nonreflecting outlet boundary
conditions when the anisotropic diffusion term is accounted for in Eq. (29),
which allows us to work with smaller sizes of the computational domain.

\section{Flow in a square-sectioned 90$^{\circ}$ pipe bend}

We now assess the performance of the nonreflecting outlet boundary conditions on a
full 3D test problem. We simulate the steady, turbulent flow in a
90$^{\circ}$ section of a curved square pipe at Re$=40000$. The numerical results
are compared with experimental measurements \cite{Sudo01} and numerical simulations
carried out with the software package FLUENT 6.2 \cite{Rup11} for the same parameters. 
The geometrical model and parameters are the same employed by Sudo et al. \cite{Sudo01}
in their experimental investigation.
\begin{figure}
\centering
\includegraphics[width=10cm]{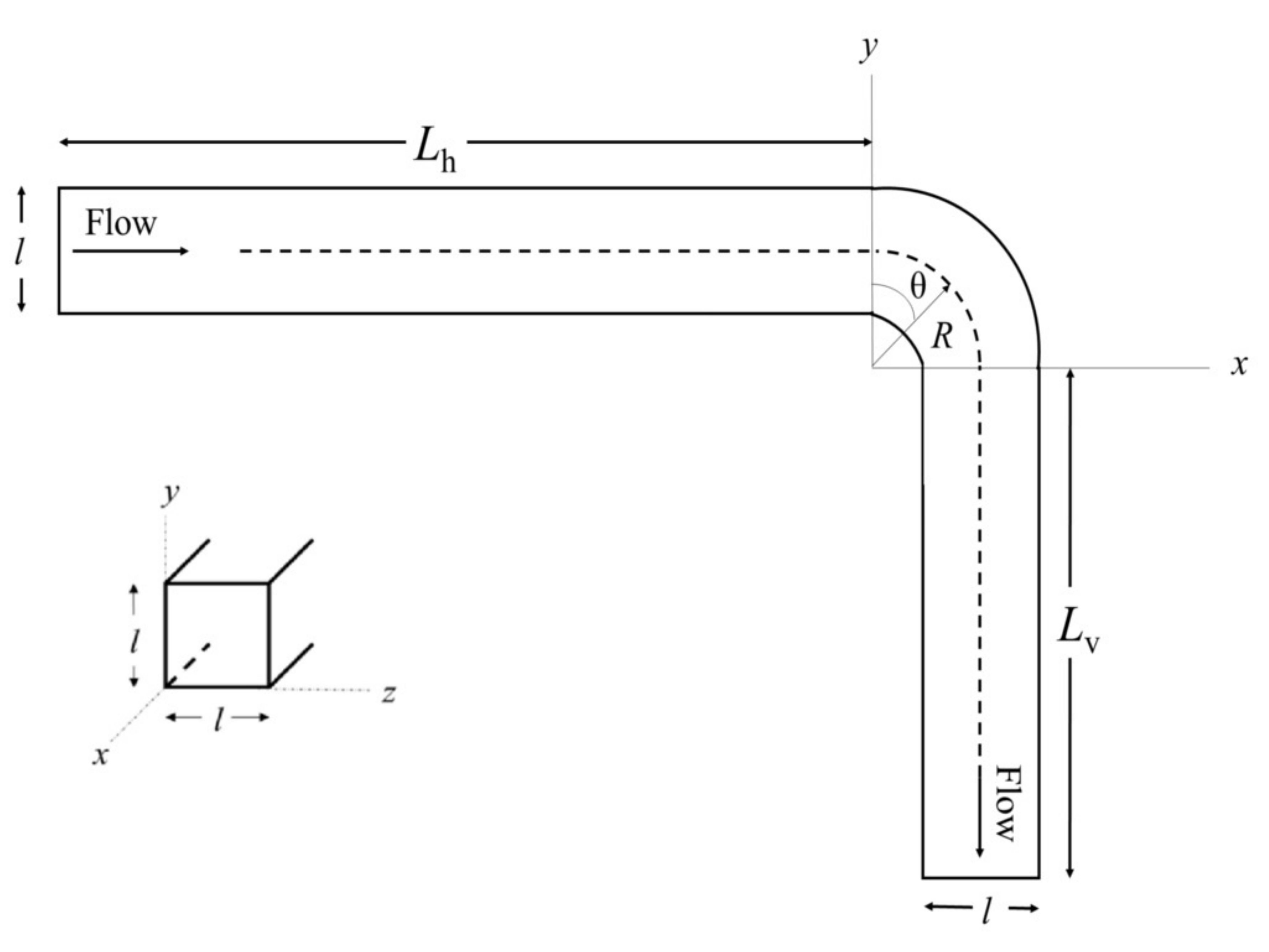}
\caption{Schematic diagrams of the pipe bend and coordinate system.}
\end{figure}

The pipe geometry is shown in Fig. 17. The pipe has a square cross-section 
measuring $l\times l=80$ mm $\times$ $80$ mm and a 90$^{\circ}$ bend of curvature radius
$R=160$ mm connected at its both ends with a horizontal straight duct upstream of
$L_{\rm h}=2$ m long and a vertical straight duct downstream of length $L_{\rm v}=1.6$ m.
At the inlet, a flat velocity profile with $v_{\rm c}=7.4$ m s$^{-1}$ is assumed in
correspondence with the experimental bulk mean velocity. With these parameters, the curvature
radius ratio is $2R/l=4$ and the Dean number is ${\rm D}={\rm Re}\sqrt{l/2R}=2\times 10^{4}$,
with Re$=v_{\rm in}l/\nu =4\times 10^{4}$, where $\nu$ is the kinematic viscosity. To
achieve a comparable spatial resolution to the simulations by Rup et al. \cite{Rup11},
we fill the pipe volume with 1.03 million particles initially at rest and uniformly
spaced in all three coordinate directions ($\Delta x=\Delta y=\Delta z=3$ mm). The
particles are given an initial smoothing length $h\approx 6.06$ mm and Eq. (4) is used
as the pressure-density relation with $c_{0}=5$ m s$^{-1}$.   
\begin{figure}
\centering
\includegraphics[width=7.3cm]{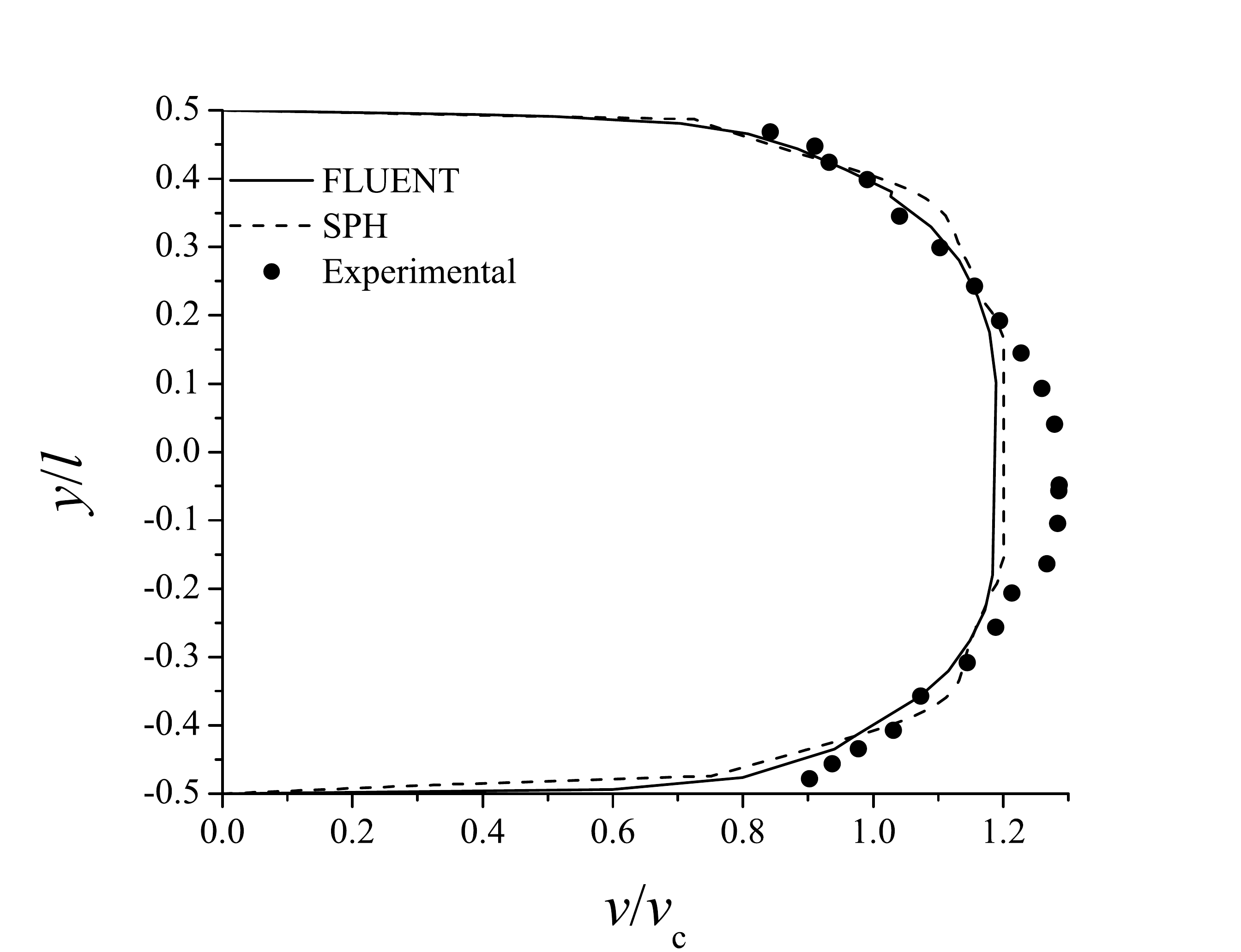}
\includegraphics[width=7.3cm]{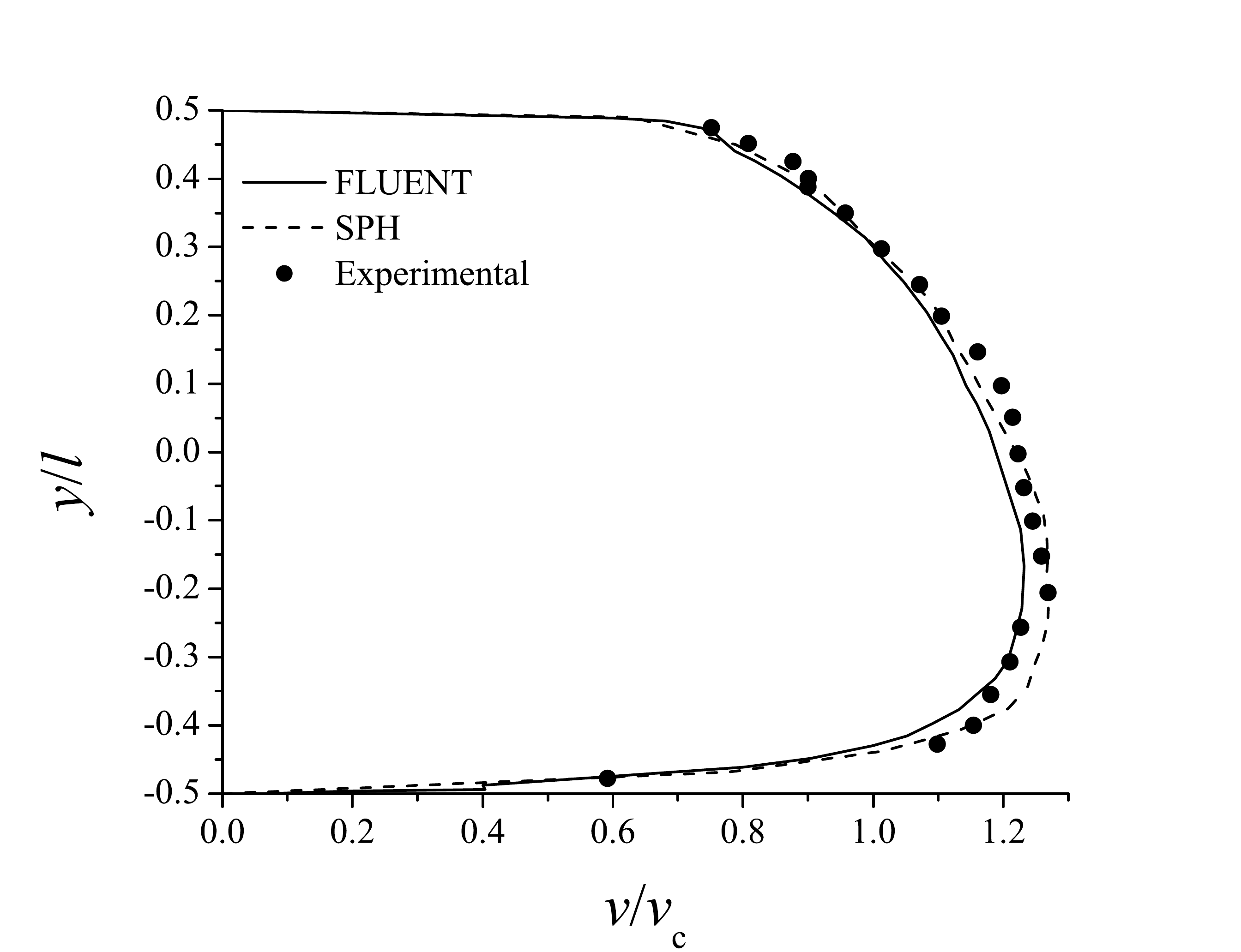}
\includegraphics[width=7.3cm]{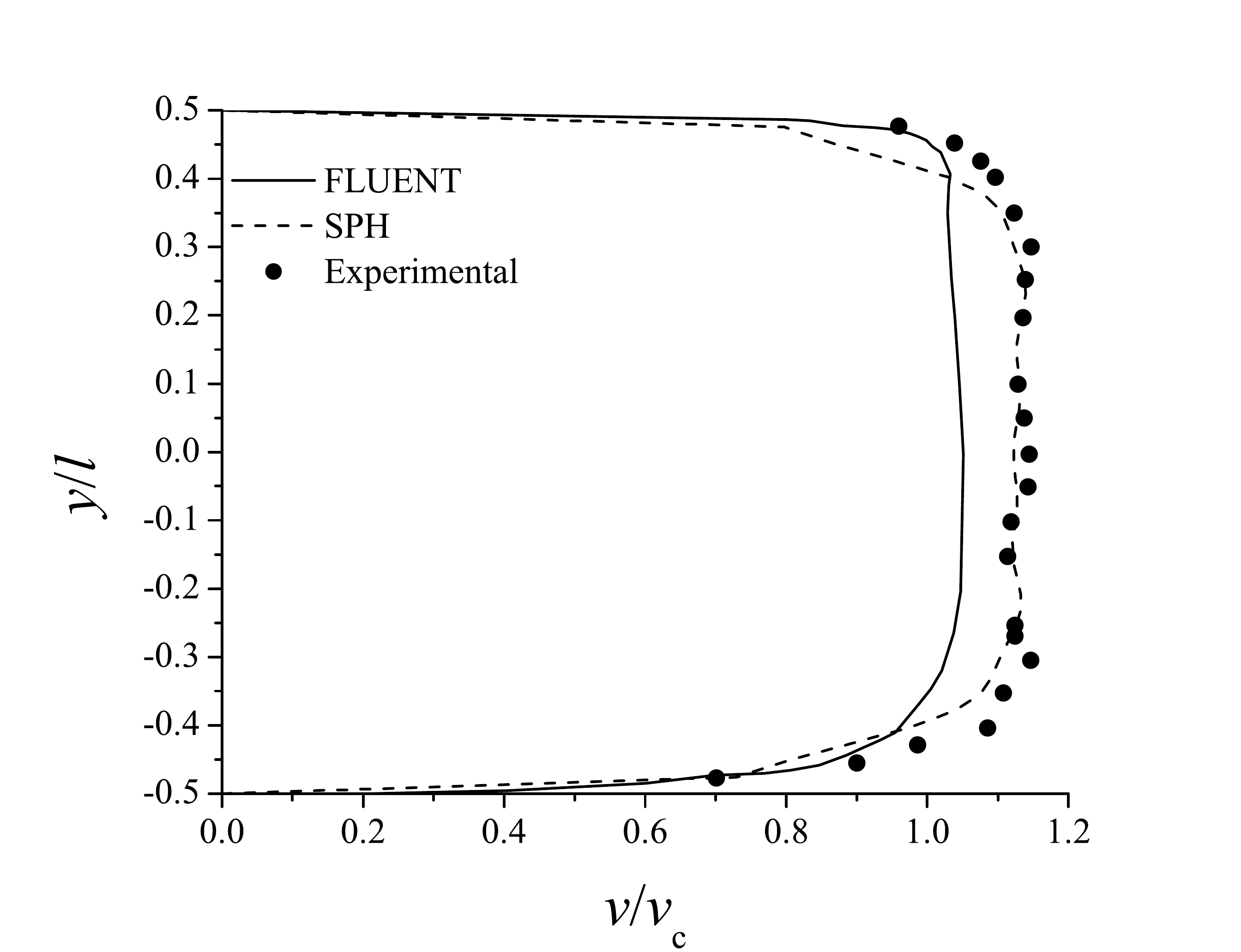}
\caption{Profiles of longitudinal mean velocity at different streamwise stations:
at $x=1.92$ m from the inlet (top frame), within the bend at $\theta =60^{\circ}$
(middle frame), and down the vertical duct at $y=0.8$ m from the bend exit (bottom 
frame). The SPH results (dashed lines) are compared with the experimental
measurements of Sudo et al. \cite{Sudo01} (dots) and the FLUENT calculations of
Rup et al. \cite{Rup11} (solid lines). The pipe cross-section and longitudinal
velocity are normalized to the hydraulic diameter $l=80$ mm and the inlet velocity
$v_{\rm c}=7.4$ m s$^{-1}$, respectively.}
\end{figure}

In order to provide direct comparison with the experimental data of Sudo et al. \cite{Sudo01}
and the numerical calculations of Rup et al. \cite{Rup11}, Fig. 18 depicts profiles of the
longitudinal mean velocity in the horizontal plane including the duct axis at three
different streamwise stations: (a) in the horizontal duct at 0.08 m from the entrance of
the bend (corresponding to $x=1.92$ m from the inlet, i.e., $z^{\prime}/d=-1$ in Sudo et al. 
\cite{Sudo01} notation), (b) within the bend at $\theta =60^{\circ}$, and (c) down the vertical
duct at $y=0.8$ m from the bend exit (i.e., $z^{\prime}/d=10$ in Sudo et al. notation). We may
see that the SPH profiles (solid lines) are in reasonably good agreement with the experimental
data (dots) and the FLUENT simulations (dashed lines). Because of the assumption of a flat
velocity profile at the inlet, the flow in the SPH and FLUENT simulations is not fully
developed at $x=1.92$ m from the inlet (top frame) and therefore the velocities around the
pipe centreline are underestimated compared to the experimental data. As the flow enters
the bend, the longitudinal velocity profile distorts as a secondary flow grows. At
$\theta =60^{\circ}$ within the bend (middle frame), the fluid flow is faster towards the
inner wall due to the larger pressure gradients there. The SPH calculation reproduces
reasonably well the asymmetric profile and closely matches the experimental and FLUENT
profiles at this station. Away from the bend exit at $y=0.8$ m (bottom frame), the
secondary flow attenuates and the vortex breaks down. At this station, the SPH simulation
reproduces very well the experimental flow velocity front, meaning that the nonreflecting
outlet boundary conditions are not influencing the flow in the elbow and along the vertical 
duct. In contrast, the FLUENT calculation underestimates the front velocity by about 10\%.
Given the good matching of the FLUENT results with the experimental measurements at 
$\theta =60^{\circ}$, the 10\% deviation in the front velocity at $y=0.8$ m from the
bend exit may be caused by some influence from the outlet boundary condition.

\section{Conclusions}

In this paper, we have described a procedure, based on Jin and Braza's \cite{Jin93}
method, for modelling nonreflecting outlet boundary conditions for incompressible
Navier-Stokes flows using the method of smoothed particle hydrodynamics (SPH). The
method, which was originally developed for two-dimensional (2D) flows, was also
generalized to three-space dimensions (3D). As it is common practice in SPH, the
method involves inflow and outflow zones of particles, which are external to the
fluid domain. A reservoir zone is designed to temporarily store particles, which is
useful in most applications where the rate of outflowing and inflowing particles is
not the same. Nonreflecting outlet boundary conditions are implemented
here by allowing the particles that leave the computational domain and enter the 
outflow zone to move according to an outgoing wave equation for the velocity field so
that feedback noises from the boundary are effectively reduced. For unsteady,
unidirectional flows, the method reduces to the well-known Orlanski wave
equation, while for steady-state flows it takes the form of a zero diffusive
boundary condition.

The performance and accuracy of the method was assessed against several 2D tests,
including the unsteady, plane Poiseuille flow, flow between inclined plates, the
Kelvin-Helmholtz instability in a channel, and flow in a constricted channel. The
performance of the method was also assessed for a 3D test problem involving the
turbulent flow in a square-sectioned 90$^{\circ}$ pipe bend. For this test, the
numerical SPH results were compared with experimental measurements and previous
numerical analysis obtained using the software package FLUENT 6.2. In general,
the results show that spurious waves incident on the outlet are effectively absorbed,
inhibiting feedback noises and allowing us to reduce the length of the computational
domain. In addition, steady-state laminar flows can be maintained stably for much
longer times compared to periodic boundary conditions. The method is stable and
has the advantage of being easily implemented for other types of incompressible
flows at low and moderate Reynolds numbers, as may be the case of flows around 
obstacles and free shear layer flows with transition towards turbulence, among
others.

\section*{Acknowledgement}
We thank the reviewers who have provided a number of comments and suggestions
that have greatly improved the style and content of the paper.
The calculations of this paper were performed using the computing facilities of
ABACUS-Cinvestav. This work was partially supported by ABACUS under CONACyT
grant EDOMEX-2011-C01-165873 and by the Departamento de Ciencias  B\'asicas
of the Universidad Aut\'onoma Metropolitana--Azcapotzalco (UAM-A) through
internal funds.

\section*{References}

\bibliography{mybibfile}

\end{document}